\documentclass[final]{cvpr}

\usepackage{times}
\usepackage{epsfig}
\usepackage{graphicx}
\usepackage{amsmath}
\usepackage{amssymb}

\usepackage{subfigure}
\usepackage{bm}
\usepackage{multirow}

\newcommand{\vhat}[1]{{\bm{\hat #1}}}
\newcommand{\vhaty}{\vhat{y}}
\newcommand{\vhatz}{\vhat{z}}
\newcommand{\hati}[2]{\hat{#1}_{#2}}
\newcommand{\hatyi}{\hati{y}{i}}

\newcommand{\talign}[2]{\multicolumn{1}{#1}{#2}}
\newcommand{\tc}[1]{\talign{c}{#1}}
\newcommand{\tcr}[1]{\talign{c|}{#1}}

\usepackage[pagebackref=false,breaklinks=true,colorlinks,bookmarks=false]{hyperref}

\def\cvprPaperID{6538} 
\def\confYear{CVPR 2021}

\begin{document}

\title{Checkerboard Context Model for Efficient Learned Image Compression}

\author{Dailan He \quad Yaoyan Zheng \quad Baocheng Sun \quad Yan Wang\thanks{Corresponding author. This work is done when Dailan He, Yaoyan Zheng and Baocheng Sun  are interns at SenseTime Research.} \quad Hongwei Qin\\
SenseTime Research\\
{\tt\small \{hedailan, zhengyaoyan, sunbaocheng, wangyan1, qinhongwei\}@sensetime.com}

}

\maketitle
\thispagestyle{empty}
\pagestyle{empty}

\begin{abstract}
For learned image compression, the autoregressive context model is proved effective in improving the rate-distortion (RD) performance. Because it helps remove spatial redundancies among latent representations. However, the decoding process must be done in a strict scan order, which breaks the parallelization.
We propose a parallelizable checkerboard context model (CCM) to solve the problem. Our two-pass checkerboard context calculation eliminates such limitations on spatial locations by re-organizing the decoding order. Speeding up the decoding process more than 40 times in our experiments, it achieves significantly improved computational efficiency with almost the same rate-distortion performance. To the best of our knowledge, this is the first exploration on parallelization-friendly spatial context model for learned image compression.
\end{abstract}


\section{Introduction}


Image compression is a vital and long-standing research topic in multimedia signal processing. Various algorithms are designed to reduce spatial, visual, and statistical redundancies to produce more compact image representations. Common image compression algorithms like JPEG~\cite{ITU1992Information}, JPEG2000~\cite{rabbani2002jpeg2000} and BPG~\cite{bellard2015bpg} all follow a general pipeline, where lossless entropy coders~\cite{marpe2003CABAC} are used after image transformations and quantization. In those non-learned image compression methods, content loss only occurs in the quantization process. The transformations involved mainly include Discrete Cosine Transformation and Wavelet Transformation, which are lossless.

In recent years, many state-of-the-art (SOTA) deep learning and computer vision techniques have been introduced to build powerful learned image compression methods. Many studies aim to establish novel image compression pipelines based on recurrent neural networks~\cite{toderici2017full}, convolutional autoencoders~\cite{balle2016end,theis2017lossy,rippel2017real,baig2017learning,balle2018variational}, or generative adversarial networks~\cite{agustsson2019generative}. Some of them~\cite{minnen2018joint,lee2018context,choi2019variable} have attained a better performance than those currently SOTA conventional compression techniques such as JPEG2000~\cite{rabbani2002jpeg2000} and BPG~\cite{bellard2015bpg} on both the peak signal-to-noise ratio (PSNR) and multi-scale structural similarity (MS-SSIM)~\cite{wang2003multiscale} distortion metrics. Of particular note is, even the intra coding of Versatile Video Coding (VVC)~\cite{bross2018versatile}, an upcoming video coding standard, has been approached by a recent learning-based method~\cite{cheng2020learned}.

\begin{figure}
\centering
\subfigure[serial $3\times 3$]{
  \includegraphics[height=1.8cm]{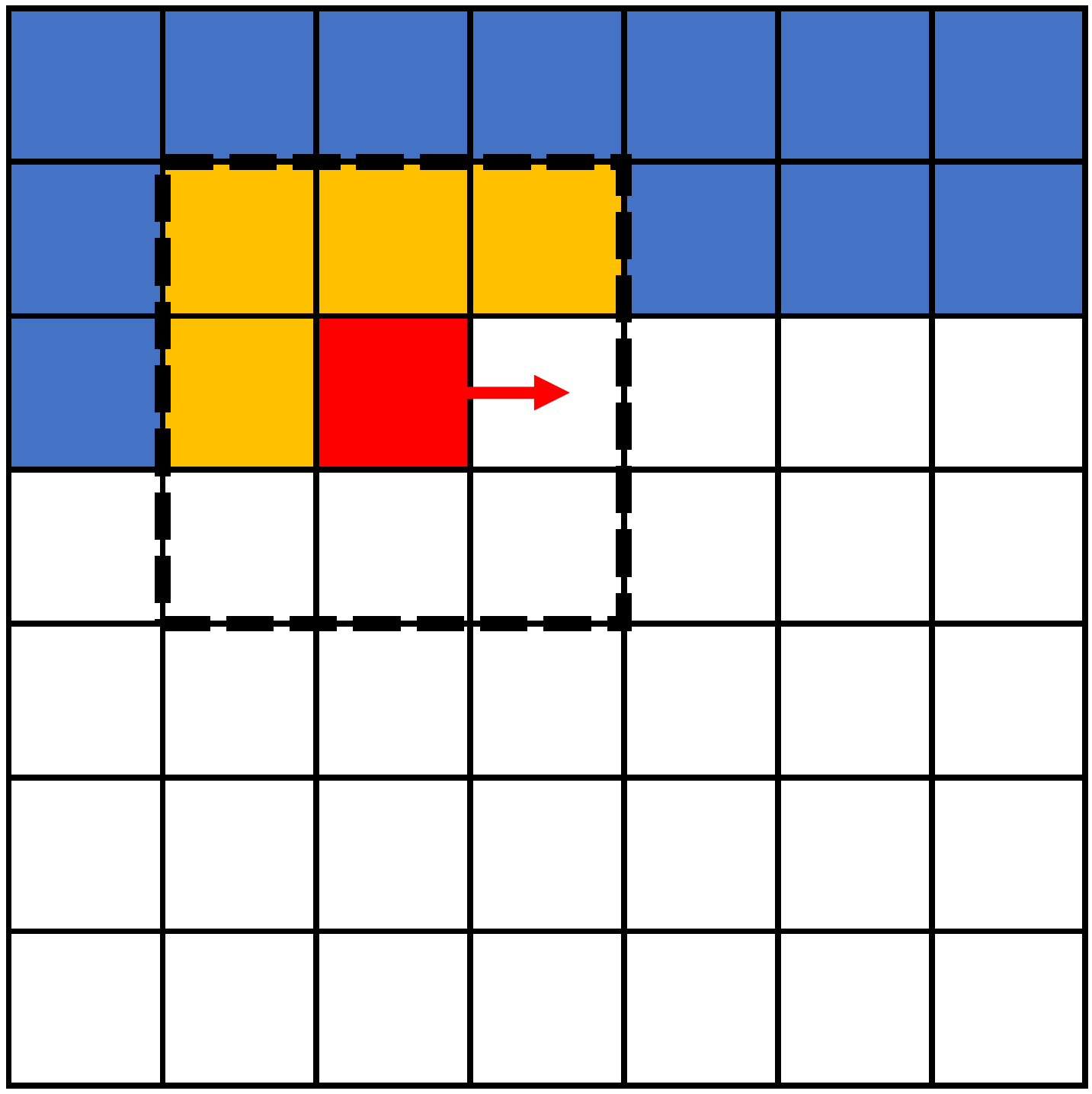}
   \label{context_models:serial3x3}
}
\subfigure[serial $5\times 5$]{
  \includegraphics[height=1.8cm]{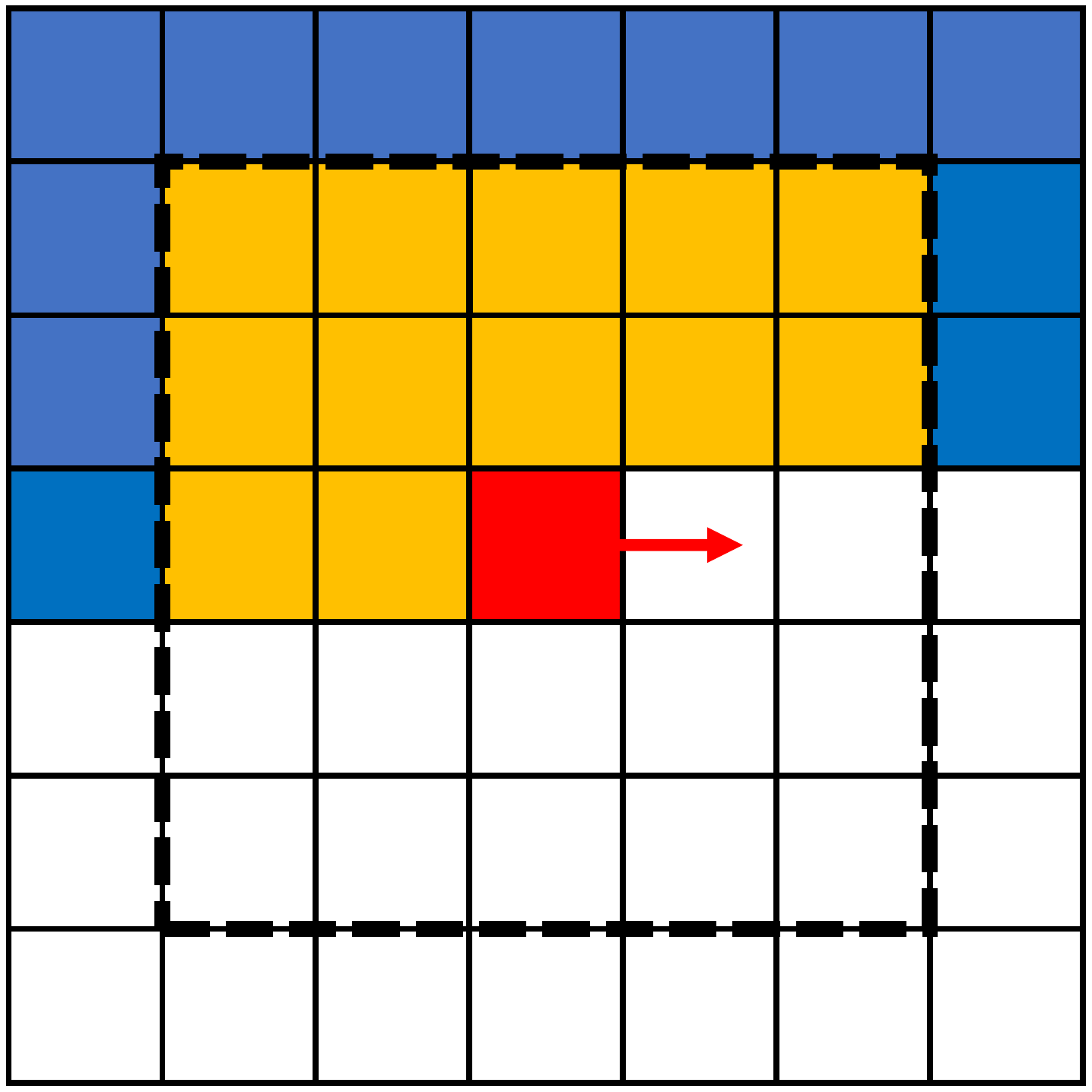}
   \label{context_models:serial5x5}
}
\subfigure[ours $3\times 3$]{
  \includegraphics[height=1.8cm]{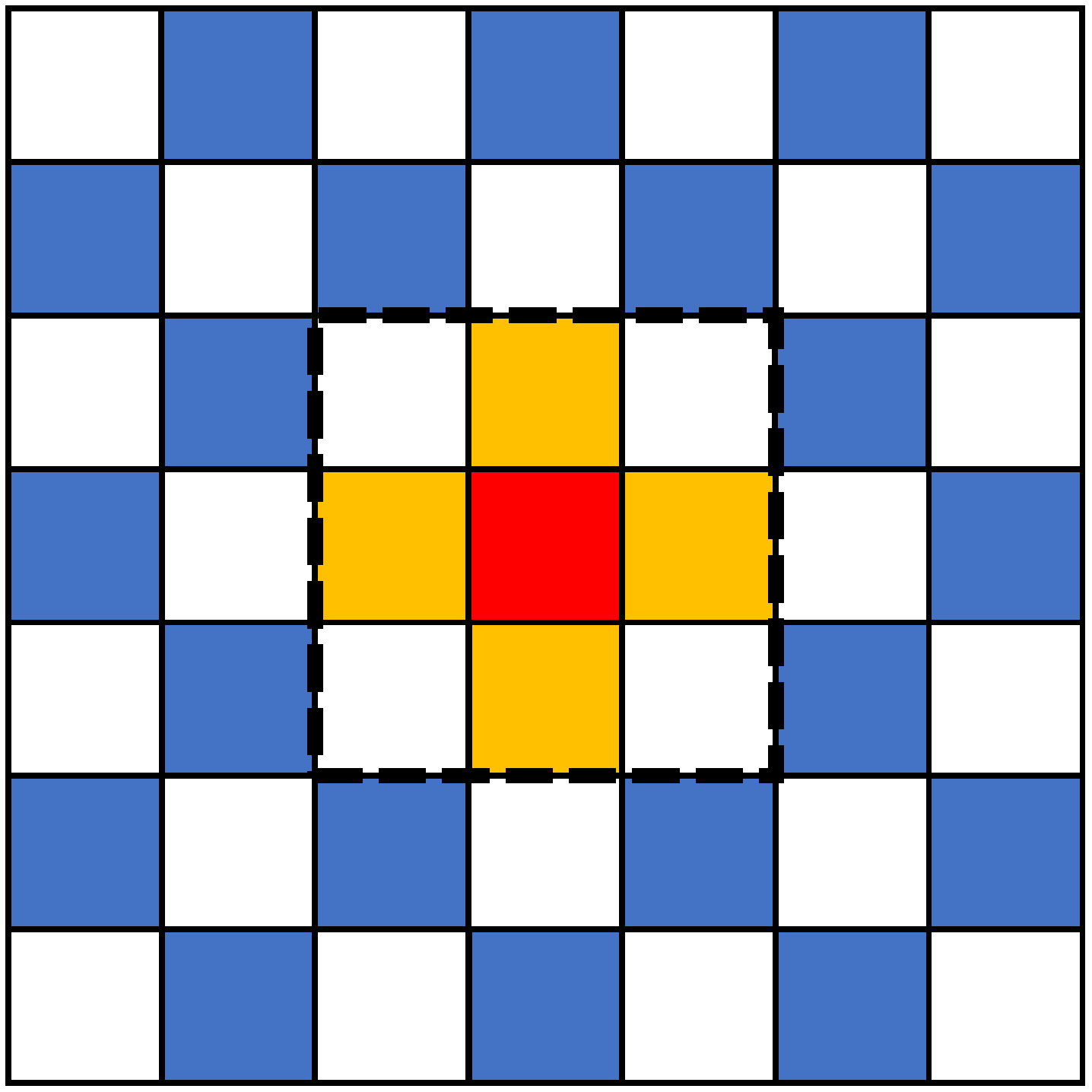}
  \label{context_models:3x3}
}
\subfigure[ours $5\times 5$]{
  \includegraphics[height=1.8cm]{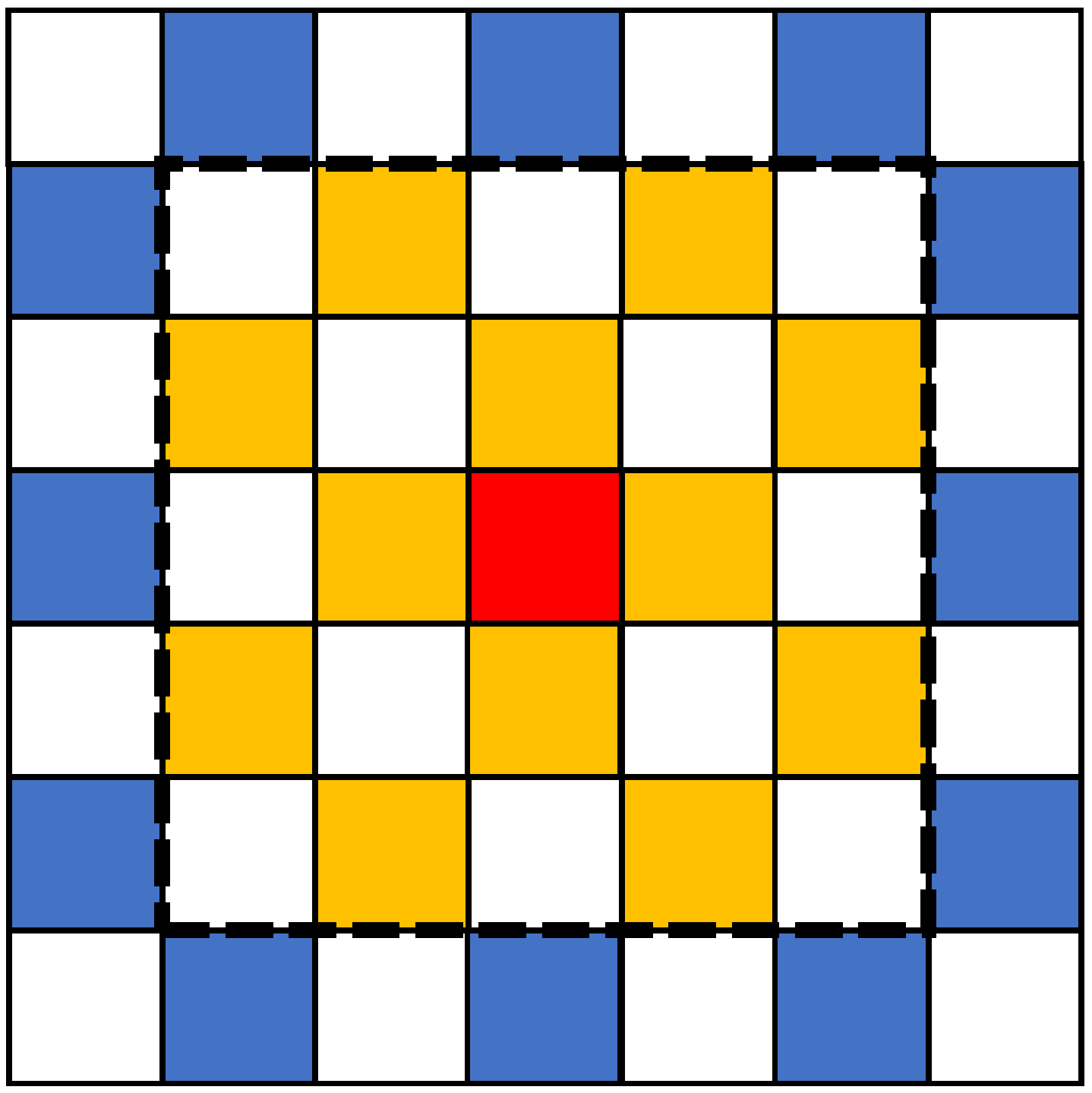}
  \label{context_models:5x5}
}
\caption{Masked convolutions modeling spatial causal context. The red blocks denote elements to en/de-code. Latents in yellow and blue locations are currently visible (all of them are visible during encoding, and those who have been decoded are visible during decoding). A context modeling can be conducted using a masked convolution which is centered at the red location and only convolves with the yellow latents. (a)(b) $3\times 3$ and $5\times 5$ instances of the widely used serial context model. This context model requires strict Z-ordered serial decoding, which limits the computational efficiency. (c)(d) Our proposed checkerboard context model with a kernel size of $3\times 3$ and $5\times 5$. 
After decoding all anchors, which are latents in blue and yellow locations, the context calculating for all non-anchors can be run in parallel.}
\label{fig:context_models}
\end{figure}

The common key of those currently most successful approaches is the entropy modeling and optimizing method, with autoencoder based structure to perform a nonlinear transform coding~\cite{goyal2001theoretical,balle2016end}. By estimating the probability distribution of latent representations, such models can minimize the entropy of these representations to be compressed, which directly correlates to the final code length using arithmetic encoding~\cite{rissanen1981universal} or range encoding~\cite{martin1979range}, and enable a differentiable form of rate-distortion (RD) optimization. Another important aspect is the introduction of hyperprior~\cite{balle2018variational}. Hyper latent is the further extracted representation, which provides side-information implicitly describing the spatial correlations in latent. Adopting hyperprior modeling allows entropy models to approximate latent distributions more precisely and benefits the overall coding performance. This method~\cite{balle2018variational} is referred to as a scale hyperprior framework in their later work~\cite{minnen2018joint}, where hyper latent is used to predict the entropy model's scale parameter.

The context model~\cite{minnen2018joint,lee2018context}, inspired by the concept of \textit{context} from traditional codecs, is used to predict the probability of unknown codes based on latents that have already been decoded (as shown in Figure~\ref{fig:context_models}(a-b)). This method is referred to as a mean-scale hyperprior framework~\cite{minnen2018joint,johnston2019computationally}, where hyper latent and context are used jointly to predict both the location (\ie mean value) and scale parameter of the entropy model.
Evaluated by previous works~\cite{minnen2018joint,lee2018context}, combining all the above-mentioned components (differentiable entropy modeling, hyper latent, and context model) can beat BPG in terms of PSNR and MS-SSIM. These context models are extended to more powerful and, of course, more computation-consuming ones by a series of later works~\cite{zhou2019multi,lee2019extended,choi2019variable,cheng2020learned}.

Though it seems promising, there are still many problems to solve before those models can be used in practice. The above-mentioned context model, which plays a crucial role in achieving SOTA performance and is adopted by most recent works, has a horribly low computational efficiency because of the lack of parallelization~\cite{minnen2018joint,johnston2019computationally}. Recent works focusing on the real-world deployment of practical neural image compression choose to omit the context model due to its inefficiency. They choose to use only the scale hyperprior framework~\cite{balle2018integer} or the mean-scale hyperprior framework but without the context model~\cite{johnston2019computationally}.  Li \etal introduce CCN~\cite{li2020efficient} for a faster context calculation with moderate parallelizability but its efficiency is still limited by image size. In order to develop a practical and more effective learning-based image codec, it is essential to investigate more efficient context models.

In this paper, we propose a novel parallelizable checkerboard context model along with a two-pass decoding method to achieve a better balance between RD performance and running efficiency.

\section{Related Work}

The most related works, including~\cite{balle2016end, balle2018variational, minnen2018joint}, which establish a powerful convolutional autoencoder framework for learned image compression, have been introduced in the introduction part.

A very recent work~\cite{minnen2020channel} proposes a channel-wise autoregressive entropy model that aims to minimize the element-level serial processing in the context model of~\cite{minnen2018joint} and achieves SOTA RD performance when combined with latent residual prediction and round-based training. Our checkerboard context is built in the spatial dimension, which provides a new solution to the serial processing problem.

There are also related works aiming to improve different aspects of the convolutional autoencoder framework, including using more complex entropy models~\cite{mentzer2018conditional,minnen2018image,lee2019hybrid,cheng2020learned}, enhancing reconstruction quality with post-processing networks~\cite{zhou2018variational}, 
enabling variable rate compression using a single model~\cite{choi2019variable,dosovitskiy2020YOTO}, using content aware compression~\cite{li2018learning}, and incorporating multi-scale or attention-based neural architectures and fancier quantization methods~\cite{mentzer2018conditional,zhou2019end,wen2019variational}.

\begin{figure}
\centering
\subfigure[autoencoder with hyperprior]{
    \includegraphics[height=2.8cm]{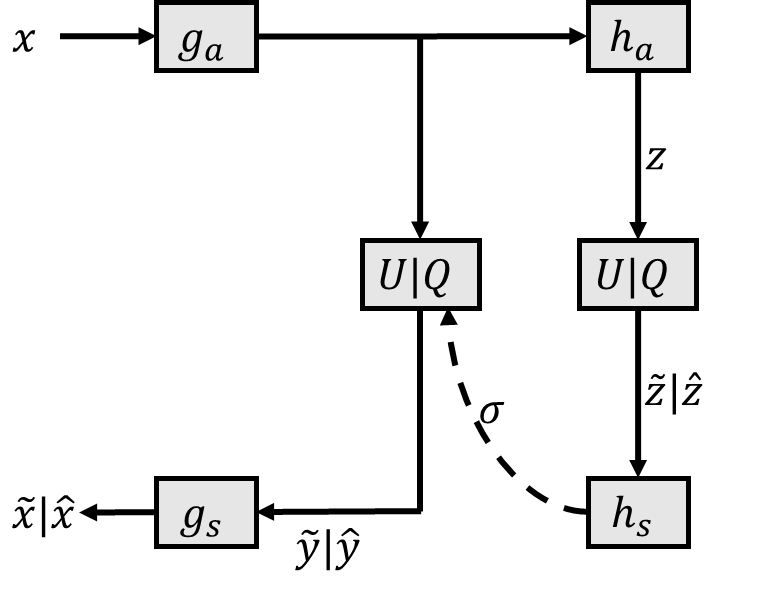}
    \label{fig:diagram:hyperprior}
}
\subfigure[extend (a) with a context model]{
    \includegraphics[height=2.8cm]{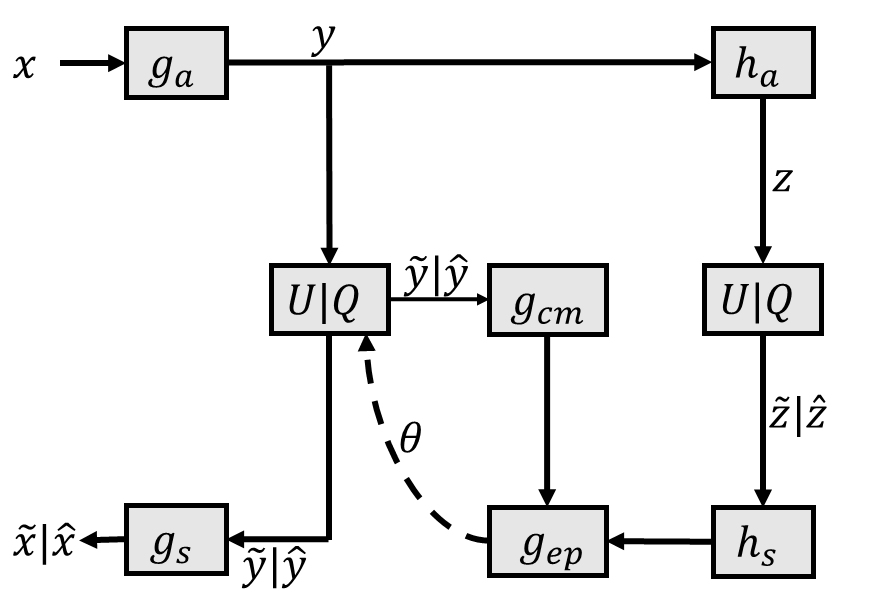}
    \label{fig:diagram:autoregression}
}
\caption{Operational diagrams: (a) Autoencoder with hyperprior. Arrows show the data flow direction. Boxes are transforms implemented by neural networks. Boxes labeled $U|Q$ is the quantization module, which performs uniform perturbation in the training stage and quantization in the inference stage. (b) Autoencoder with both hyperprior and context model.}
\label{fig:diagram}
\end{figure}

\section{Preliminary}
\subsection{Variational Image Compression with Hyperprior}
The diagram for the scale hyperprior framework~\cite{balle2018variational} is given in Figure~\ref{fig:diagram:hyperprior}. $g_a, g_s, h_a, h_s$ are nonlinear transforms implemented by neural networks. $\bm x$ is the original image, $\bm y=g_a(\bm x)$ and $\vhat y = Q(\bm y) $ are latent representations before and after quantization, $\vhat x = g_s(\vhat y)$ is the reconstructed image. $\bm z=h_a(\bm y)$ and $\vhat z = Q(\bm z)$ are the hyper latent before and after quantization. $\vhat z$ is used as side information to estimate the scale parameter $\bm \sigma$ for the entropy model of latent $\vhat y$. During training, the quantization operation is approximated by adding uniform noise, producing differentiable variables $\bm {\tilde y}$, $\bm {\tilde z}$ and $\bm {\tilde x}$. Hereinafter we always use $\vhat x$, $\vhat y$ and $\vhat z$ to represent $\bm {\tilde x}| \vhat x$, $\bm {\tilde y}| \vhat y$ and $\bm {\tilde z}| \vhat z$ for simplicity.
The tradeoff between rate and distortion or the loss function can be written as:
\begin{equation}
\begin{aligned}
R + \lambda \cdot D &=  \mathbb{E}_{\bm x \sim p_{\bm x}}[-\log_2 p_{\vhat y|\vhat z}(\vhat y | \vhat z) -\log_2 p_{\vhat z}(\vhat z) ] \\
&+ \lambda \cdot \mathbb{E}_{\bm x \sim p_{\bm x}} [d(\bm x, \vhat x)]
\end{aligned}
\end{equation}
where bit rate of latent $\vhat y$ and hyper latent $\vhat z$ is approximated by estimated entropy, $\lambda$ controls the bit rate(\ie larger $\lambda$ for larger rate and better reconstruction quality), $d(\bm x, \vhat x)$ is the distortion term, usually using MSE or MS-SSIM.

With scale hyperprior~\cite{balle2018variational}, the probability of latents $\vhat y$ can be modeled by a conditional Gaussian scale mixture (GSM) model:
\begin{equation}
\begin{aligned}
p_{\vhat y|\vhat z}(\hatyi | \vhat z) = \left[ \mathcal{N}(\mu_i, \sigma_i^2) * \mathcal{U}(-\frac{1}{2}, \frac{1}{2}) \right] (\hatyi)
\end{aligned}
\end{equation}
\begin{equation}
\begin{aligned}
p_{\vhat y|\vhat z}(\vhaty | \vhat z) = \prod_i p_{\vhat y|\vhat z}(\hatyi | \vhat z)
\end{aligned}
\end{equation}
where the location parameter $\mu_i$ is assumed to be zero and the scale parameter $\sigma_i$ is the i-th element of $\bm \sigma=h_s(\vhat z)$ for each code in $\vhat y$ given the hyperprior. The probability of the hyper latent $\vhat z$ can be modeled using a non-parametric fully factorized density model~\cite{balle2018variational}. 

\subsection{Autoregressive Context}
In the mean-scale hyperprior framework~\cite{minnen2018joint}, an additional module called context model is added to boost the RD performance. Figure~\ref{fig:diagram:autoregression} shows the entire structure consisting of autoencoder ($g_a, g_s$), hyper autoencoder ($h_a, h_s$), context model ($g_{cm}$), and a parameter inference network ($g_{ep}$) which estimates the location and scale parameters $\bm{\Phi} = (\bm \mu, \bm \sigma)$ of the entropy model for latent $\vhat y$. Let $h_s(\vhatz)$ denote the hyperprior feature and $g_{cm}(\vhat y_{<i})$ denote the context feature, the parameter prediction for $i$-th representation $\hat y_i$ is
\begin{align}
\bm \Phi_i = (\mu_i, \sigma_i)= g_{ep}\big(h_s(\vhatz), g_{cm}(\vhat y_{<i}) \big)
\label{eq:gep-vanilla-context-model}
\end{align}
where $\vhat y_{<i}$ means the causal context (\ie some nearby visible latents of latent $\hat y_i$). This type of context model can be implemented using masked convolutions. Given the $k\times k$ binary mask $\bm M$ and convolutional weights $\bm W$, the masked convolution with kernel size of $k\times k$ on input $\bm x$ can be performed in a reparameterization form:
\begin{equation}
g_{cm}(\bm x) = (\bm M \odot \bm W)\bm x + \bm{b}
\label{eq:masked-conv-gcm}
\end{equation}
where $\odot$ is the Hadamard operator and $\bm b$ is a bias term. Since $\bm{M}$ describes the context modeling pattern, by manually set different mask $\bm M$ various context referring schemes are obtained. In previous works, usually the mask shown in Figure~\ref{context_models:serial3x3} is adopted as $\bm M$ to implement a left-top reference which requires strict Z-ordered serial decoding because only already decoded latents are visible.

In~\cite{cheng2020learned}, the entropy model is improved by using a discretized  K-component Gaussian mixture model (GMM):
\begin{equation}
\begin{aligned}
p_{\vhaty | \vhatz}(\hatyi|\vhatz) = \sum_{0<k<K} \pi_i^{(k)} \left[ \mathcal{N}(\mu^{(k)}_i, \sigma^{2(k)}_i) * \mathcal{U}(-\frac{1}{2}, \frac{1}{2})\right]
(\hatyi)
\end{aligned}
\end{equation}
where K groups of entropy parameters $(\bm \pi^{(k)}, \bm \mu^{(k)}, \bm \sigma^{(k)})$ are calculated by $g_{ep}$. Combined with the autogressive context model, \cite{cheng2020learned} is the first to achieve comparable PSNR with VVC.

Following previous works~\cite{minnen2018joint,cheng2020learned}, we do not apply context model to the hyper latent $\vhat z$.

\section{Parallel Context Modeling}

Though the previous context model, called \textbf{serial} context model by us and shown in Figure~\ref{context_models:serial3x3}, has a limitation on computational efficiency, most of the SOTA methods still rely on it~\cite{minnen2018joint, cheng2020learned, lee2019hybrid}. 
Therefore, improvement is highly required. We firstly analyze the context model using a random-mask model, then we propose a novel parallel context model with checkerboard shaped masked convolution as a replacement to the existing serial context model.

\subsection{Random-Mask Model: Test Arbitrary Masks} \label{section: Random-Mask Model}

\begin{figure}
    \centering
    \includegraphics[width=6cm]{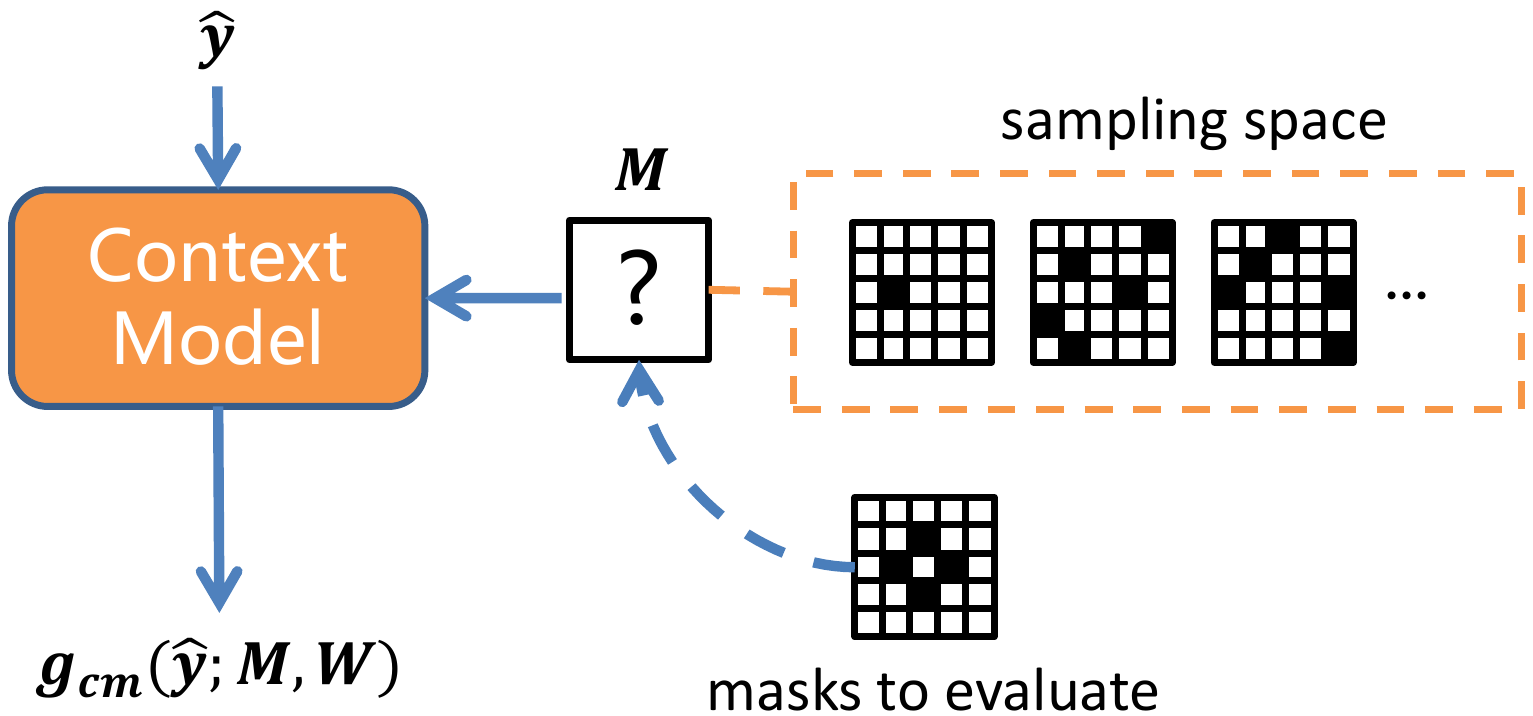}
    \caption{Context model in proposed random-mask model. Mask $\bm{M}$ is randomly generated during training (orange dashed box) and replaced by a fixed mask (blue dashed line) during evaluation.}
    \label{fig:random-mask-model}
\end{figure}

The context model can be seen as a convolution with weight $\bm{W}$ conditioned on the binary mask $\bm{M}$, as described in eq.~\ref{eq:masked-conv-gcm} where the mask describes the pattern of context modeling. 
To understand the mechanism of context models and explore better modeling patterns, we propose a random-mask toy model to which arbitrary mask can be fed as the context modeling pattern.

The random-mask model is adapted from the above-mentioned autoregressive model~\cite{minnen2018joint} with the same en/de-coders and $g_{ep}$. The serial context model with a manually designed mask is replaced by a convolution conditioned on randomly generated masks during training. As shown in Figure~\ref{fig:random-mask-model}, we generate random sampled $5\times 5$ masks $\bm{M}$ and compute Hadamard product of $\bm{M}$ and non-masked convolution weights $\bm{W}$ to obtain the masked convolution weights $\bm{M}\odot \bm{W}$ in every iteration of the training stage. After each time of backward propagation, the weights $\bm{W}$ will be updated guided by a random mask $\bm{M}$, which implicitly establishes a supernet consisting of all context models conditioned on $5\times 5$ masks. Therefore, weights are shared among context models using different masks so that the trained random-mask model can be used to evaluate arbitrary masks during inference.
After training, to test the performance of a particular mask pattern, we simply feed that mask to the context model as $\bm{M}$, and then the random-mask model has a context model with a fixed mask. 

This random-mask model is suitable to measure the ability to save bit rate brought by various context modeling patterns. The same latent $\vhaty$ and reconstructed image $\vhat x$ are shared (because of the weight sharing) so that the same distortion is shared among context modeling patterns, enabling the direct comparison of bit rates. Let $R_0$ denote the non-reference bits per pixel (BPP)  when context modeling is disabled by inputting a full-zero mask to the random-mask model (corresponds to the context-free mean-scale hyperprior baseline~\cite{minnen2018joint}). We further quantitate the ability of rate saving as a rate saving ratio:  
\begin{equation}
\eta(\bm M) = \frac{R_0 - R_{\bm M}}{R_0} \times 100\%
\end{equation}
where $R_{\bm M}$ denotes the BPP feeding mask $\bm M$ to the random-mask model. In the following section, we use the trained random-mask model to analyze context modeling patterns represented by masks $\bm{M}$ by calculating their rate saving ratio $\eta(\bm{M})$.

\subsection{How Distance Influences Rate Saving}

\begin{figure}
    \centering
    \subfigure[]{
    \includegraphics[height=4cm]{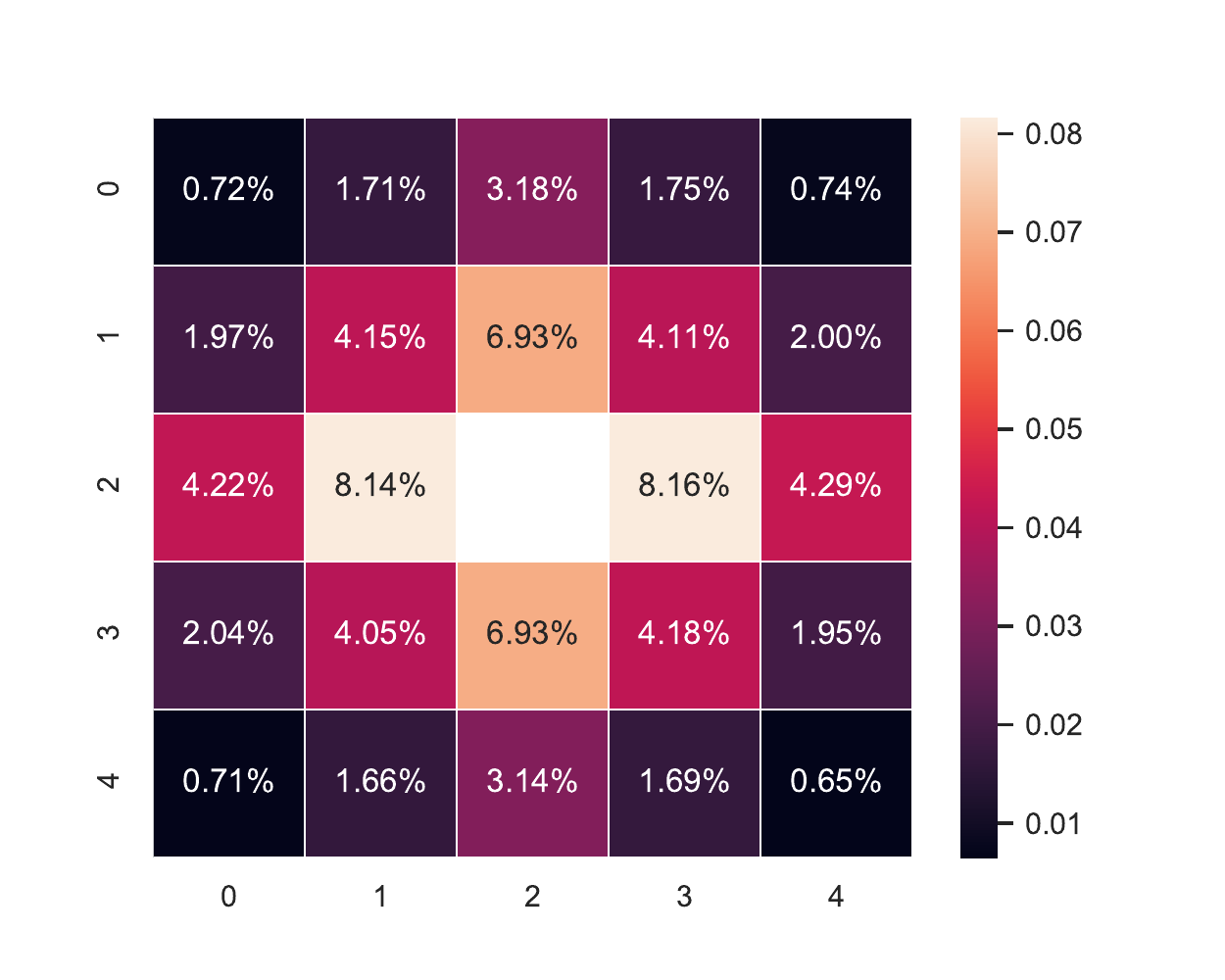}
    \label{fig:random-mask:single-refer:result}
    }
    \subfigure[]{
    \includegraphics[height=3cm]{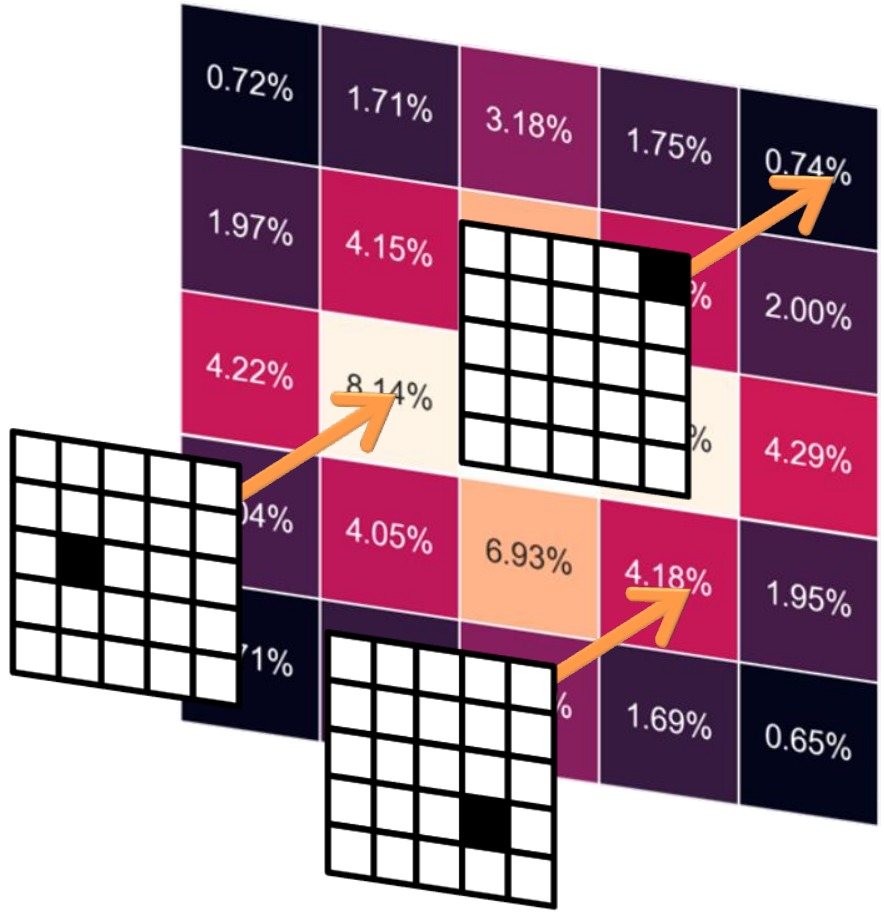}
    \label{fig:random-mask:single-refer:diagram}
    }
    
    \caption{(a) Rate saving ratios of single-reference masks tested on Kodak images~\cite{kodak}. (b) For each single-reference mask, only one location is set to 1 (black) and others are set to 0 (white).}
    \label{fig:random-mask:single-refer}
\end{figure}

The serial context model saves rate by referring to already decoded neighboring latents to estimate the entropy of current decoding latent more precisely. For such a model with a $5\times 5$ mask (Figure~\ref{context_models:serial5x5}), latents on 12 locations at the left-top of the central location are referred to in the estimation. We find that latents on nearer locations contribute to rate saving much more significantly by calculating the $\eta$ of 24 different single-reference masks (each mask has only one location set to 1 and others set to 0, as shown in Figure~\ref{fig:random-mask:single-refer:diagram}) based on the above-mentioned random-mask model. The result is shown in Figure~\ref{fig:random-mask:single-refer:result}. It is obvious that closer neighboring latents reduce much more bit rate, and neighbors with a distance of more than 2 elements influence the bit saving negligibly.

To dig deeper, we analyze how the context models help save rate. With context modeling, the entropy model estimates decoding latents $\vhat y_i$ conditioned on those visible already decoded latents $\vhat y_{j<i}$. Viewing those latents on different locations as samples from correlated virtual coding sources $\hat Y_i$ and $\hat Y_{j<i}$, the mutual information $I(\hat Y_i;\hat Y_{j<i})$ between them is recognized by context model and partially removed from bits used to encode $\vhat y_i$. According to the Slepian-Wolf coding theory~\cite{slepian1973noiseless}, the optimal bit rate:
$$\hat R = H(\hat Y_i, \hat Y_{j<i}) = H(\hat Y_i) + H(\hat Y_{j<i}) - I(\hat Y_i;\hat Y_{j<i})$$
is theoretically reachable. Therefore, provided that the training is sufficient, a context modeling pattern referring to latents with more mutual information $I(\hat Y_i;\hat Y_{j<i})$, or simply saying, more causal relationship, saves more bit rate.

On the other hand, spatial redundancy is an important basis of image and video compression. Strong self-correlation exists in digital images, and adjacent pixels are likely to have a stronger causal relationship.
Empirically, convolution outputs partially keep such redundancy because of locality~\cite{balle2018variational}, even when using scale hyperprior~\cite{minnen2018joint,cheng2020learned}. So the latents still retain a similar redundancy.

That tells how the distance between referred latents and decoding latents influences the bit saving of context modeling in learned image compression. Nearer latents have a stronger causal relationship, and more mutual information can be re-calculated during decoding instead of being stored and occupying the bit rate. Also, it helps explain why modeling the context using masked convolution with larger kernels or a stack of $5\times 5$ convolutions (with a much larger receptive field) unexpectedly does damage to the RD-performance, as reported in previous work~\cite{minnen2018joint}. Referring to further neighbors that carry nearly no mutual information helps little but increases the risk of overfitting. 

This implies that a context modeling pattern referring to more close neighbors is more likely to save more bit rate than presently adopted serial context models. We then establish our parallel context model based on this motivation.

\subsection{Parallel Decoding with Checkerboard Context}

\begin{figure*}
    \centering
    \includegraphics[width=13cm]{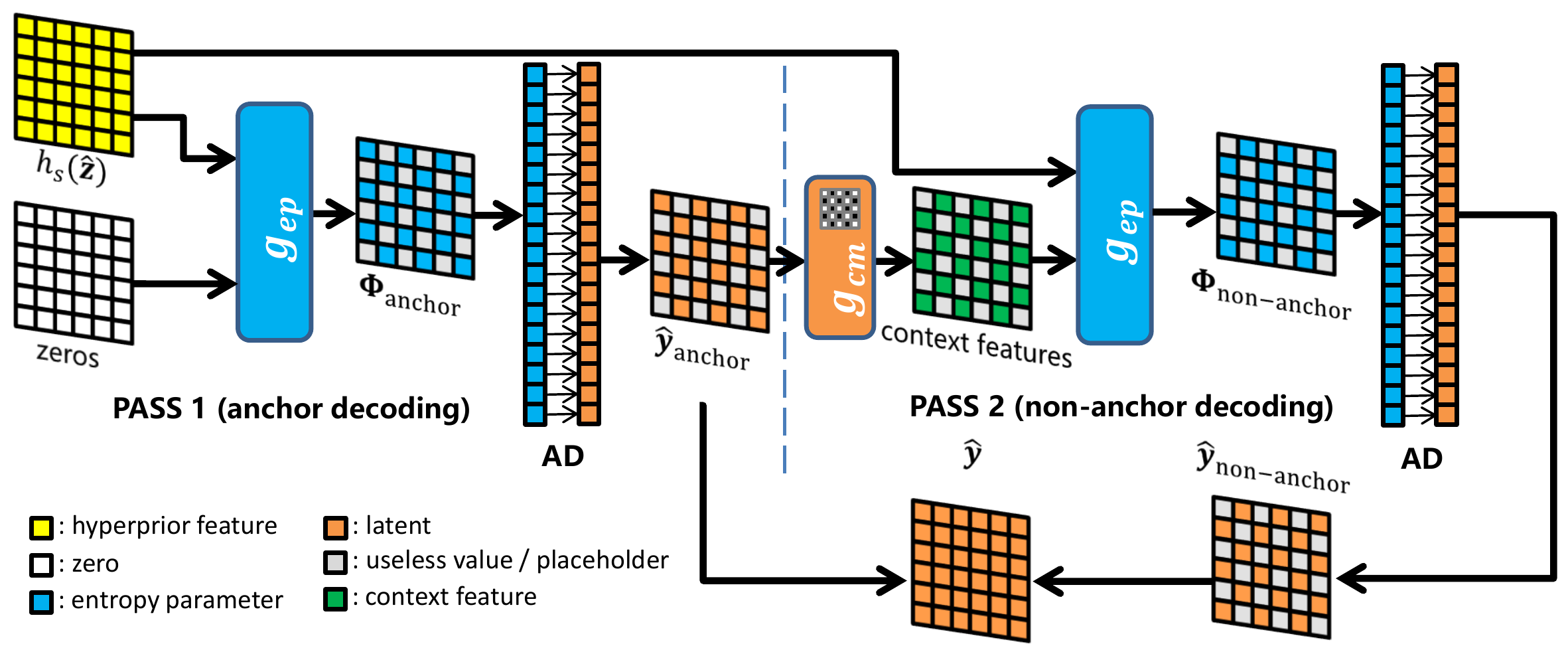}
    \caption{Illustration of the proposed two-pass decoding. $g_{cm}$ is the context model with a checkerboard mask and $g_{ep}$ is the parameter network. Entropy parameters $\bm{\Phi}_{\rm{anchor}}$ and $\bm{\Phi}_{\rm{non-anchor}}$ are estimated in turn. To decode the latents (orange blocks) from bitstream (not given in the diagram), useless values (gray blocks) in estimated entropy parameters are removed and the remained ones (blue blocks) are flattened and input into arithmetic decoder (AD). For further implementation details, please refer to the supplementary material. }
    \label{fig:two-pass-dec}
\end{figure*}

Figure~\ref{context_models:3x3} shows a checkerboard shaped context model. Referring to four nearest neighbours, it outperforms both $3\times 3$ and $5\times 5$ serial context model in our further experiments on random-mask model (we will discuss this experiment in section~\ref{section: toy-details} and Table~\ref{tab:bpp_estimation}, here we bring forward its conclusion to motivate the proposal of checkerboard shaped context models). Then we further extend it to a general form with arbitrary kernel size (for an instance of $5\times 5$ kernel, see Figure~\ref{context_models:5x5}). Though it is impossible to apply it on the whole latent feature map (or the adjacent latents will depend on each other during decoding), it is helpful to building a computationally efficient parallel context model without introducing apparent RD performance loss compared with the present serial context model.

\newcommand{\vhatyanchor}[0]{\vhaty_{\rm{anchor}}}
\newcommand{\vhatynonanchor}[0]{\vhaty_{\rm{non-anchor}}}

To develop our parallel decoding approach, we only encode/decode half of the latents (white and red ones in Figure~\ref{context_models:3x3} and Figure~\ref{context_models:5x5}) using checkerboard shaped context and hyperprior. The coding of the other half of latents, which we call \textbf{anchors}, only depends on the hyperprior. To implement these two sets of rules, we set the context feature of all anchors zero and adapt the calculation of entropy parameters $\bm{\Phi}$ in eq.~\ref{eq:gep-vanilla-context-model} to a spatial location conditioned form:
\begin{equation}
\bm{\Phi}_i=\left\{
\begin{aligned}
& {g_{ep}\left(h_s(\vhatz), \bm{0}\right)}_i, & \hatyi\in \vhatyanchor \\
& {g_{ep}\left(h_s(\vhatz), g_{cm}(\vhatyanchor; \bm{M}\odot \bm{W})\right)}_i,  & \rm{otherwise} \\
\end{aligned}
\right.
\label{eq:phi-calculation}
\end{equation}
where $g_{cm}$ is the masked convolution as described in eq.~\ref{eq:masked-conv-gcm} conditioned on a checkerboard-shaped mask $\bm{M}$.
Its input $\vhatyanchor$ is the set of anchors and $i$ is the index for the i-th element $\hatyi$ in latent $\vhat y$. For approaches using mean-scale Gaussian entropy models, the entropy parameter $\Phi = (\bm{\mu}, \bm{\sigma})$, and for methods adopting GMM $\Phi$ consists of K groups of $\bm{\pi}^{(k)}$, $\bm{\mu}^{(k)}$ and $\bm{\sigma}^{(k)}$.

When anchors are visible, the context features of all non-anchors can be calculated in parallel by a masked convolution. Anchors' decoding is also run in parallel, so the entropy parameter calculation in eq.~\ref{eq:gep-vanilla-context-model} for decoding can be performed in two passes, which is much more efficient than the serial context model. 

\subsubsection{Encoding Latents in One Pass}

\newcommand{\vhatymix}[0]{\vhaty_{\rm{half}}}
\newcommand{\bhalf}[0]{\bm{b}_{\rm{half}}}

Here we reformulate eq.~\ref{eq:phi-calculation} to illustrate that the encoding process can be done within one pass. By one pass we mean entropy parameters $\bm{\Phi}$ of all the latents are obtained in parallel without element-wise sequential calculation.
Let  $ \vhatymix$ denotes the latents with all non-anchors set to zero:
\begin{equation}
(\vhatymix)_i=\left\{
\begin{aligned}
& \hatyi, & \hatyi\in \vhatyanchor \\
& 0,  & \rm{otherwise} \\
\end{aligned}
\right.
\end{equation}
Since now the binary mask $\bm{M}$ is checkerboard-shaped, if input $\vhatymix$ into $g_{cm}$, according to eq.~\ref{eq:masked-conv-gcm} we have:
\begin{equation}
g_{cm}(\vhatymix)_i=\left\{
\begin{aligned}
& b_{\hatyi}, & \hatyi\in \vhatyanchor \\
& g_{cm}(\vhatyanchor)_i,  & \rm{otherwise} \\
\end{aligned}
\right.
\end{equation}
where $b_{\hatyi} \in \bm b$ is the corresponding bias term added to $\hatyi$. Similar to $\vhatymix$, let $\bhalf$ denote a feature map with all elements on anchor locations set to $\bm{b}$ and the others set to zeros.
Because $g_{ep}$ is usually implemented as a point-wise transform consisting of a stack of $1\times 1$ convolutions~\cite{minnen2018joint, cheng2020learned}, we can re-write eq.~\ref{eq:phi-calculation} to:
\begin{equation}
\label{eq:phi-one-pass}
\bm{\Phi}=g_{ep}\left(h_s(\vhatz), g_{cm}(\vhatymix)-\bhalf\right)
\end{equation}
During encoding (and training) all latents in $\vhaty$ are visible, so we can simply generate $\vhatymix$ from $\vhaty$ by setting its non-anchors to zero and then calculate all entropy parameters $\bm{\Phi}$ in parallel with only one pass of context model and parameter network. Finally, we flatten and rearrange $\vhaty$ and $\bm{\Phi}$ and then encode anchors and non-anchors into the bitstream in turn.  We will discuss cheap and parallel ways to get $\vhatymix$ and $\bhalf$ in the supplementary material. Hence, only an element-wise subtraction in eq.~\ref{eq:phi-one-pass} is newly introduced to the encoding process for using the checkerboard context model, which won't slow down the encoding.

\subsubsection{Decoding Latents in Two Passes}

At the beginning of decoding, the hyper latent $\bm{\hat z}$ can be decompressed directly by an arithmetic decoder (AD) from the bitstream using code probability $p_{\bm{\hat z}}$. Then the hyperprior feature $h_s(\bm{\hat z})$ is calculated. After that, there are two decoding passes to obtain whole latent $\bm{\hat y}$, as shown in Figure~\ref{fig:two-pass-dec}.

In the first decoding pass, entropy parameters of anchors $\bm{\Phi}_{\rm{anchor}}$ are calculated with the context feature set to zero according to eq.~\ref{eq:phi-calculation}.  Then the conditional probability $p_{\vhat y| \vhat z}$ of anchors is determined by these entropy parameters. Now the anchors, half of all the latent elements in $\bm{\hat y}$, can be decoded by AD. Then decoded anchors become visible for decoding non-anchors in the next pass. 

In the second decoding pass, the context feature for non-anchors can be calculated using proposed checkerboard model and entropy parameters of them are calculated from concatenated hyperprior feature and context feature. Then AD can decompress the remained half of $\bm{\hat y}$ (those non-anchors) from the bitstream and we obtain whole latent $\bm{\hat y}$. Finally, we get reconstructed image $\bm{\hat x} = g_s(\bm{\hat y})$.

 \subsubsection{Structure and Analysis} \label{section: structure}

\begin{figure*}
\centering
\includegraphics[width=16cm]{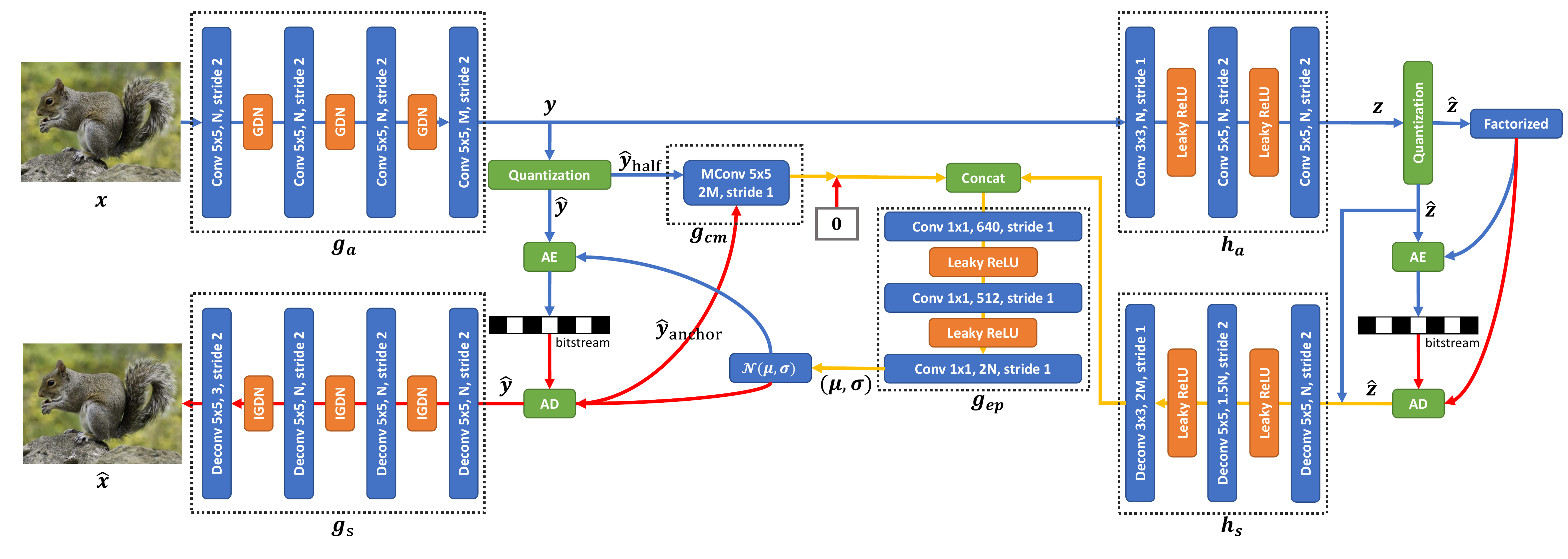}
\caption{
Previously proposed framework of learned compression model using context based autoregressive entropy model~\cite{minnen2018joint}. We replace the serial context model with the proposed checkerboard convolution. Blue and red lines denote encoding and decoding data flow respectively. Processes shared by both encoding and decoding are represented by yellow lines. 
}
\label{fig:pipeline}
\end{figure*}

As an example, Figure~\ref{fig:pipeline} shows how we adapt the structure of the autoregressive model~\cite{minnen2018joint} by replacing its serial context model with proposed checkerboard context model. We don't modify other components of the autoregressive model for fair comparison. 

For convolution kernel size of $k$, $\frac{k^2} 2$ neighbors evenly distribute around the center (as is shown in Figure~\ref{context_models:3x3}) and contribute to the context for each non-anchor spatial location. Though only half of the latents refers to their neighbours now, a checkerboard context model can extract more causal information from decoded neighbours to save more bit rate. This compensates the potential BPP increase when compressing anchors without using any context in our proposed method.

Compared with the serial context model~\cite{minnen2018joint} which requires $H\times W$ sequential steps to decode a $H\times W \times M$ latent feature map, our proposed parallel model allows a constant step of 2 to decode such latents, where $H\times W \times \frac M 2$ latent representations are processed in parallel. Considering that large-size images with more than one million pixels are produced and shared frequently nowadays, this parallelizable checkerboard context model brings significant improvement on practicality.

\begin{figure}
    \centering
    \includegraphics[width=7.5cm]{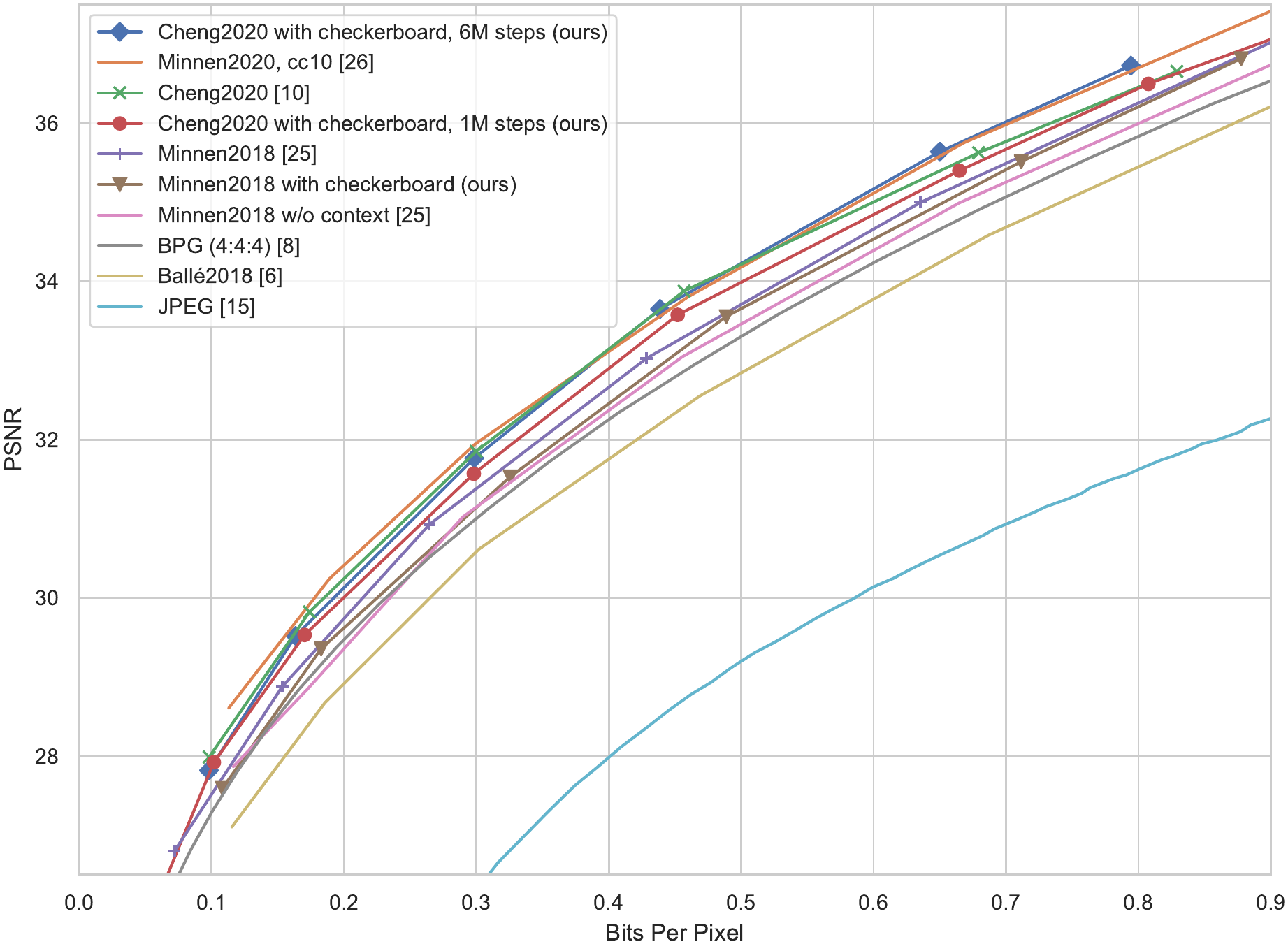}
    \caption{
    RD curves of various learned or manually designed image codecs. The results are averaged on Kodak. Except for two models with our proposed context model, all data are reported by prior works. All learned models are optimized for MSE. For MS-SSIM optimization results please refer to the supplementary material.
    }
    \label{fig:ablation-context-model}
\end{figure}

\begin{figure}
    \centering
    \includegraphics[width=7.5cm]{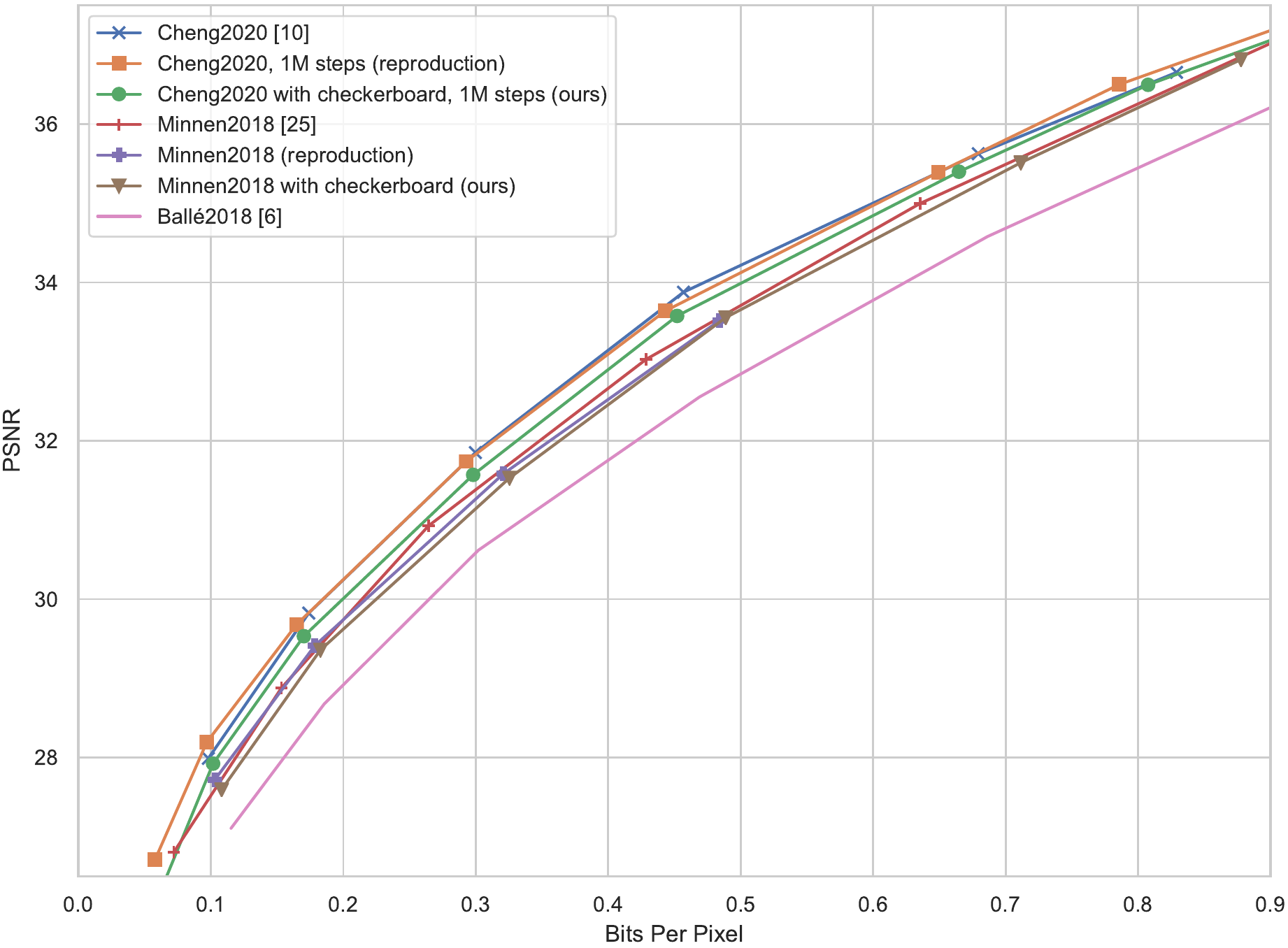}
    \caption{
    Comparison of parallel and serial context models on Minnen2018 and Cheng2020. Curves marked as \textit{ours} and \textit{reproduction} are trained on our dataset using above specified settings, while the rest are reported results.
    }
    \label{fig:ablation-context-model-}
\end{figure}

\section{Experiments}

We implement representative previous works and our proposed methods in PyTorch~\cite{paszke2019pytorch}. We choose the largest 8000 images from the ImageNet~\cite{deng2009imagenet} validation set as our training data, where each image has more than one million pixels. Following previous works~\cite{balle2016end,balle2018variational}, we add random uniform noise to each of them and then downsample all the images.  We use Kodak dataset~\cite{kodak} and Tecnick dataset~\cite{tecnick2014TESTIMAGES} as our test set.

During training, before fed into models all input images are randomly cropped to $256 \times 256$ patches. All models are trained for 2000 epochs (\ie 2M steps) with a batch-size of 8 and learning rate of $10^{-4}$ if not specified. Sadam optimization~\cite{balle2018efficient} is adopted on each convolution layer for training stability.

\subsection{Toy Experiments on Random-Mask Model} \label{section: toy-details}

We adapt the architecture of Minnen2018~\cite{minnen2018joint} by replacing its context model by a $5 \times 5$ random-mask convolution to obtain the random-mask model mentioned in section~\ref{section: Random-Mask Model}. Here we further discuss its details. To generate a binary mask, each location of the mask is sampled from $\{0, 1\}$ in equiprobability. During training, new random masks are generated in every iteration. To train such a model with random-mask context, we set $\lambda=0.01$ and $N=M=256$ for MSE optimization. On the trained model we perform two toy experiments:
\begin{itemize}
    \item {
    \textbf{Single Reference Mask for Causal Estimation}. As above mentioned, we estimate the causal relationship between latent pairs by using $5\times 5$ masks with only one location set to $1$ and calculating their $\eta$ on Kodak.
    Figure~\ref{fig:random-mask:single-refer:result} shows the result which implies that the mutual information between latents decays quickly as the spatial distance between them increases.
    }
    \item {
    \textbf{Performance of Various Masks}. We test several masks using the random-mask model and post a brief report in Table~\ref{tab:bpp_estimation}. It indicates that simply increasing the number of referred neighbours (\ie $K_{ref}$ in the table) doesn't always help save bit rate, because the pattern referring to 4 adjacent neighbours (the row \textit{checkerboard $3\times 3$}) performs much better than the pattern referring to 12 left-top neighbours (the row \textit{serial $5\times 5$}). Also, even referring to all 8 surrounding neighbours cannot outperform \textit{checkerboard $3\times 3$}. However, masks referring to more adjacent neighbours (the last three rows) always perform better. This further proves that closer neighbours play much more important roles in context modeling. Motivated by this experiment, we establish our parallel context model based on \textit{checkerboard $5\times 5$}.
    }
\end{itemize}

\begin{table}[]
    \centering
    \begin{tabular}{c|c|c|r}
         description & $K_{ref}$ & $R_{\bm{M}}$ & \tc{$\eta(\bm{M})$}\\
         \hline
          non-reference ($R_0$) &0 & 0.4332 & 0.0\%\\
         serial $3\times3$ (Fig.~\ref{context_models:serial3x3})& 4 & 0.3928 & 9.3\% \\
         serial $5\times5$ (Fig.~\ref{context_models:serial5x5})& 12& 0.3817 & 11.9\%     \\
        checkerboard $3\times 3$ (Fig.~\ref{context_models:3x3}) & 4 & 0.3651 & 15.7\% \\
          checkerboard $5\times 5$ (Fig.~\ref{context_models:5x5}) & 12 & 0.3648 & 15.8\% \\
         all neighbours in $3\times 3$ & 8 & 0.3649 &  15.8\%\\
    \end{tabular}
    \caption{Various masks tested with the random-mask model. $K_{ref}$ denotes the number of referred neighbours (\ie number of ones in the mask). $R_{\bm M}$ and $\eta(\bm{M})$ denote the BPP and rate saving ratio when feeding mask $\bm M$ to the random-mask model. The row \textit{all neighbours in $3 \times 3$} refers to the mask where all 8 neighbours surrounding the center are set to 1. Note that the random-mask model is not designed as an actual codec, during decoding we can use entropy model parameters obtained in the encoding process, which is not allowed in actual decoder.
    }
    \label{tab:bpp_estimation}
\end{table}

\begin{table*}[]
    \centering
    \begin{tabular}{c|r|r|r|r|r|r }
         \multirow{2}{*}{architecture (N=192)} & \multirow{2}{*}{Ball\'{e}2018} & \multicolumn{3}{c|}{Minnen2018}  & \multicolumn{2}{c}{Cheng2020}  \\
          \cline{3-7}
          & & \tcr{w/o context} & \tcr{serial} & \tcr{parallel(ours)} & \tcr{serial} & \tc{parallel(ours)}  \\
         \hline
         \hline
         Kodak   ($768\times 512$)               & 26.34 & 26.41 &    1323.66       &   29.66 & 1395.35       &   75.23  \\
         \hline
          Tecnick      ($1200\times 1200$)            & 83.28 & 86.31 &   4977.98  &  95.46 & 5296.16 &   259.37 \\

    \end{tabular}
    \caption{Total decoding time averaged on Kodak and Tecnick (ms).  Feature map size of each Kodak image is $48 \times 32 \times M$ and feature map size of each Tecnick image is  $75 \times 75 \times M$.  Ball\'{e}2018 is the earlier context-free hyperprior model~\cite{balle2018variational} while Minnen2018 and Cheng2020 represent the serial autoregressive structure~\cite{minnen2018joint} and its Gaussian Mixture Model (GMM) and attention involved adaption~\cite{cheng2020learned} respectively. For models marked as \textit{serial} the serial context models are adopted and for \textit{parallel} ones the proposed checkerboard context is used.
    }
    \label{tab:latency}
\end{table*}

\begin{table}[]
    \centering
    \begin{tabular}{c|r|r }
         Minnen2018 (N=192)& 
          \tcr{serial} & \tc{parallel(ours)}  \\
         \hline
         \hline
         hyper synthesis         & 1.26    &   1.42    \\
         parameter calculation  &  1302.42 &	4.75 \\
         latent synthesis     &   20.98   &	23.49   \\
         \hline
         total & 1323.66 & 29.66

    \end{tabular}
    \caption{Running speed of Minnen2018's each decoding process (ms). In \textit{hyper synthesis} and \textit{latent synthesis}, the decoders $h_s$ and $g_s$ are invoked respectively. In \textit{parameter calculation}, context features and entropy parameters are calculated by $g_{cm}$ and $g_{ep}$. }
    \label{tab:latency:tecnick}
\end{table}

\subsection{Checkerboard Context Model V.S. Serial Context Model}

 We evaluate our checkerboard context model and the parallel decoding method based on two previous architectures using serial context: Minnen2018~\cite{minnen2018joint} and Cheng2020~\cite{cheng2020learned}. As is explained in section~\ref{section: structure} and shown in Figure~\ref{fig:pipeline} using Minnen2018 as an example, we do not change any other architectures except replacing the serial context model by the proposed one to ensure fair comparisons.

To compare the performance of different context models, we implement these two baseline models, and for each baseline we train its adaption with parallel context model under the same settings. The detailed experimental settings are:
\begin{itemize}
    \item \textbf{Cheng2020}. We train Cheng2020 following their reported settings, \ie K = 3,  $\lambda$ = \{0.0016, 0.0032, 0.0075, 0.015, 0.03, 0.045\}, N = M = 128 for the three lower $\lambda$ and N = M = 192 for the rest\footnote{The hyper-parameter M is not introduced in the original paper. Here we let M denote number of output channel of the encoder $g_a$'s last layer.}. For reproduction we train 1M steps, and we find that training up to 6M steps can lead to an even better RD performance especially at high bit rates. During training, after 3M steps we decay the learning rate from $10^{-4}$  to $5\times 10^{-5}$. 
    \item \textbf{Minnen2018}. As the exact setting was not reported, we choose to use the same $\lambda$ setting as Cheng2020. We follow the suggestion on reproduction from authors\footnote{\texttt{https://groups.google.com/g/tensorflow- \\ compression/c/LQtTAo6l26U/m/cD4ZzmJUAgAJ}} to use N = 128 and M = 192 for small $\lambda$ and N = 192 and M = 320 for big $\lambda$ and consider $\lambda  \geq 0.015$ (corresponding BPP $> 0.6$) as big ones. For all models we train 6M steps and the learning rate decays to $5\times 10^{-5}$ after 3M steps.

\end{itemize}

\subsubsection{Decoding Speed}

We evaluate inference latency of both serial and parallel models on Nvidia TITAN XP: Table~\ref{tab:latency} and Table~\ref{tab:latency:tecnick}. Notice that for proposed parallel models, the context model and the parameter network are invoked twice in the parameter calculation process. For serial models, context model and parameter network are called $H \times W$ times for decoding a $H\times W \times M$ feature map. For each time we crop the visible neighbours to $5\times 5$ patches then feed them into the serial context model for a fast implementation~\cite{zhou2019multi} to pursue a fair comparison. It is obvious that a serial context model becomes a bottleneck even when decoding relatively small images. Meanwhile, the proposed two-pass decoding model has a much faster running speed on parallel devices, making spatial context based image compression models more practical.

\subsubsection{Rate-Distortion Performance}

We train and compare proposed checkerboard context model and the serial model based on Minnen2018 and Cheng2020 with settings as mentioned above. Figure~\ref{fig:ablation-context-model-} shows their RD-curves evaluated on Kodak (the ones marked as \textit{with checkerboard}). Compared with original models, using such a parallel context model only slightly reduces RD performance on Kodak. However, it still outperforms the context-free hyperprior baseline (Ball\'e2018 in the figure) significantly (compared with Ball\'e2018, BDBR $-17.0\%$/$-27.4\%$ for Minnen2018/Cheng2020 
using our checkerboard context model). Since it removes the limitation of computational efficiency, we think the quality loss is acceptable.

For completeness, we also compared our proposed models with several previous learned or conventional codecs including a recently proposed channel conditioned model~\cite{minnen2020channel}, a SOTA architecture without using spatial context. See Figure~\ref{fig:ablation-context-model}, the channel conditioned model performs better than Cheng2020 with a parallel context model at lower bit rate but slightly worse at higher bit rate. As discussed by the author, a potential combination of spatial context and channel-wise adaption is promising to further improve RD performance of present approaches, since only the spatial-wise adaption can help remove spatial redundancy from source images. However, investigating the proper way of combining the two different type of adaptive modeling is beyond the scope of this paper as our topic is to speed up the spatial context model. To our understanding, the proposed parallel and efficient spatial context model is an important basis for future researches on the multi-dimension context modeling techniques, or the running speed will be inevitably limited by the low-efficiency serial context model.

We have also tested all above-mentioned models on Tecnick dataset and come to the same conclusion that the proposed checkerboard context model can be used as an efficient replacement to the serial context. For more RD curves and results please refer to the supplementary material.

\section{Discussion}

Serial context models for learned image compression are computationally inefficient. We propose a parallel context model based on the checkerboard-shaped convolution and develop a two-pass parallel decoding scheme. Compared with the serial context model, it allows decoding to be implemented in a highly parallel manner. After applying it to two representative context model involved baselines, we prove that it greatly speeds up the decoding process on parallel devices while keeps a competitive compression performance. Also, our proposed approach does not require any changes in model structure or model capacity, so it is almost a drop-in replacement for the original widely used serial context model. Therefore, the proposed approach significantly improves the potential performance of SOTA learned image compression techniques with spatial context models. 

{\small
\bibliographystyle{ieee_fullname}
\bibliography{egbib}
}

\end{document}


\title{Supplementary Material
}


\maketitle

\section{Detailed Running Speed}

\begin{table*}[]
    \centering
    \begin{tabular}{c|c||r|rr|r||r|r}
        \multicolumn{2}{c||}{\multirow{2}{*}{architecture (N=192)}} & hyper & \tc{parameter} & \tcr{\multirow{2}{*}{\%}} & latent &  \tcr{total} & \tc{speed} \\
         \multicolumn{2}{c||}{} & synthesis & \tc{calculation} & & synthesis & \tcr{(ms)} & \tc{(Mpps)}\\
         \hline \hline
    \multicolumn{8}{c}{Kodak, image size: $768\times 512$, feature size: $48\times 32 \times M$} \\
         \hline \hline
         \multicolumn{2}{c||}{Ball\'e2018} 
         & 1.30 & - & - & 25.04 &   26.34 & 14.93\\
         \hline
         \multirow{3}{*}{Minnen2018} 
         & w/o context
         & 1.32 & 2.23 & 8.4\% & 22.85   & 26.41 & 14.89\\
         & serial 
         & 1.26 & 1302.42 & 98.4\% & 20.98 & 1323.66 & 0.30\\
         & parallel (ours)
         & 1.42 & 4.75 & 16.0\% & 23.49 & 29.66 & 13.26\\
         \hline
         \multirow{2}{*}{Cheng2020}
         & serial 
         & 1.82 & 1325.32 & 95.0\% & 68.21 & 1395.35 & 0.28\\
         & parallel (ours)
         & 1.72 & 4.37 & 5.8\% & 69.14 & 75.23 & 5.23\\
         \hline \hline
    \multicolumn{8}{c}{Tecnick, image size: $1200\times 1200$, feature size: $75\times 75 \times M$} \\
         \hline \hline
         \multicolumn{2}{c||}{Ball\'e2018} 
        & 2.22 & - & - & 81.06 & 83.28 & 17.29\\
         \hline
         \multirow{3}{*}{Minnen2018} 
         & w/o context
        & 4.01 & 1.30 & 1.5\%  & 81.00 & 86.31 & 16.68\\
         & serial 
        & 3.78 & 4891.43 & 98.3\% & 82.77 & 4977.98 & 0.29\\
         & parallel (ours)
        & 4.27 & 8.36 & 8.8\% & 82.83 & 95.46 & 15.08\\
         \hline
         \multirow{2}{*}{Cheng2020}
         & serial 
        & 3.40 & 5044.93 & 95.3\% & 247.83 & 5296.16 & 0.27\\
         & parallel (ours)
        & 3.09 & 10.98 & 4.2\% & 245.30  & 259.37 & 5.55\\
    \end{tabular}
    \caption{Running time of each decoding process (unit: microsecond) on GPU. Meaning of table headers is the same as which is in Table~2 and Table~3 in the main body. The column \textit{speed} shows million-pixels-per-second (Mpps) of each model,  which represents how fast a specific architecture can decode images from bitstream.}
    \label{tab:detailed-decoding-time}
\end{table*}

\begin{table*}[]
    \centering
    \begin{tabular}{c|c||r|rr|r||r|r}
        \multicolumn{2}{c||}{\multirow{2}{*}{architecture (N=192)}} & hyper & \tc{parameter} & \tcr{\multirow{2}{*}{\%}} & latent &  \tcr{total} & \tc{speed} \\
         \multicolumn{2}{c||}{} & synthesis & \tc{calculation} & & synthesis & \tcr{(s)} & \tc{(Kpps)}\\
         \hline \hline
    \multicolumn{8}{c}{Kodak, image size: $768\times 512$, feature size: $48\times 32 \times M$} \\
         \hline \hline
         \multicolumn{2}{c||}{Ball\'e2018} 
         & 0.059  & - & - &  1.460 &    1.519 & 258.89 \\
         \hline
         \multirow{3}{*}{Minnen2018} 
         & w/o context
         & 0.118  & 0.009  &  0.5\% &  1.720   & 1.848  & 212.79 \\
         & serial 
         & 0.115 & 20.402 &  99.4\% & 1.875 & 20.518 & 19.23\\
         & parallel (ours)
         & 0.113 & 0.083 & 3.9\% & 1.937 & 2.132 & 184.42\\
         \hline
         \multirow{2}{*}{Cheng2020}
         & serial 
         &  0.108 &  22.107 &  91.7\% & 1.901  & 24.116 & 16.30 \\
         & parallel (ours)
         &  0.095 & 0.086  &  4.6\% &  1.678
 &  1.859 & 211.49
 \\
    \end{tabular}
    \caption{Running time of each decoding process (unit: second) on CPU. }
    \label{tab:detailed-decoding-time-cpu}
\end{table*}

We give detailed report on decoding time on our PyTorch implementation. Extending Table~2 and Table~3 in the main body, latency for each process averaged on Kodak and Tecnick is shown in Table~\ref{tab:detailed-decoding-time}. It is apparent that for all serial models, calculating context features and entropy parameters is the bottleneck, which takes more than 95\% decoding time. By adopting the proposed parallel context model, Minnen2018's average decoding speed increases 44.6 times on Kodak (small images) and 52.1 times on Tecnick (large images). Meanwhile, Cheng2020 also speeds up 18.5 and 20.4 times tested on the two datasets due to a parallel context model.

We also test above mentioned models on CPUs as a reference (see Table~\ref{tab:detailed-decoding-time-cpu}). Though we mainly aim to improve the decoding efficiency on parallel devices, our parallel context model can still perform well on CPUs because of a sufficient use of matrix librarys \eg Intel MKL, which is very encouraging. We measure the speeds by running our PyTorch implementation on Intel Xeon E5-2620.

\section{More RD Curves}

For completeness, we evaluate all models on Tecnick. Figure~\ref{fig:rd-tecnick} shows the results. Compared with a channel conditioned solution Minnen2020, Cheng2020 with proposed checkerboard context model performs slightly worse in the low bit rate but better in the high bit rate. This perhaps because GMM cannot estimate the density precisely when using less coding channels (N = 128 in our case), as described in the original Cheng2020 paper. On the other hand, Minnen2018 with checkerboard context model even slightly outperforms original Minnen2018 baseline in higher bit rates. It further proves that a checkerboard-shaped convolution can extract more causal relationship and helps save more rate, especially in larger images which have more smooth area with more spatial redundancy.

We also optimize our models for MS-SSIM with $D = 1 - {\rm{MSSSIM}} (\bm{x}, \vhat x)$ and test all models with MS-SSIM. To train models for MS-SSIM optimization, we set $\lambda$ to $\{1, 3, 16\}$ for low bit rate models (N = 128) and $\{40, 120, 360\}$ for high rate models (N = 192). Figure~\ref{fig:rd-ms} shows the results tested on Kodak.

\section{Cheap Operations for Parallel En/De-coding}

\subsection{Multiplexer: Generating $\bm{\hat y}_{\rm{half}}$ from $\bm{\hat y}$}

\newcommand{\yhalf}[0]{\vhaty_{\rm{half}}}
\newcommand{\bhalf}[0]{\bm{b}_{\rm{half}}}

As described in the main body, we generate $\yhalf$ and $\bhalf$ to perform a one-pass encoding. To implement it we further define a multiplexer operation $\mux$:
\begin{equation*}
\mux(\bm{\alpha}, \bm{\beta})_i=\left\{
\begin{aligned}
& \alpha_i,  & \hat{y}_i\in \bm{\hat y}_{\rm{anchor}} \\
& \beta_i,  & otherwise \\
\end{aligned}
\right.
\end{equation*}
where inputs $\bm{\alpha}$ and $\bm{\beta}$ are feature maps with the same sizes $H\times W\times M$ as $\vhaty$. So we can simply calculate $\yhalf$ and $\bhalf$ with below formula:
\begin{equation*}
    \yhalf = \mux(\vhaty, \bm{0})
\end{equation*}
\begin{equation*}
    \bhalf = \mux(\bm{b}_{\rm{map}}, \bm{0})
\end{equation*}
where $\bm{b}_{\rm{map}} \in \mathbb{R}^{H\times W\times M}$ is the feature map with every $H\times W$ locations filled with the $M$-element bias vector $\bm{b}$.

Since we do not want the newly introduced multiplexer operation slow down the  encoding process, it must be elaborately arranged as a series of light operations like slicing or viewing operators in PyTorch. We provide two of practical ways to implement the multiplexer.

\subsubsection{Using slice-assign operations}

For frameworks such as PyTorch and TensorFlow supporting slice-and-assign operators, we simply use such operators with a slicing step of 2 to mix the anchors and non-anchors input into the multiplexer. It could be a piece of code like:

\begin{lstlisting}
# mix = MUX(alpha, beta)
mix = clone(alpha)
mix[..., 1::2, 0::2] = beta[..., 1::2, 0::2]
mix[..., 0::2, 1::2] = beta[..., 0::2, 1::2]
\end{lstlisting}

\subsubsection{Using slicing and concatenating}
\label{sec:using-slice-and-concat}

\begin{figure}
    \centering
    \includegraphics[height=4cm]{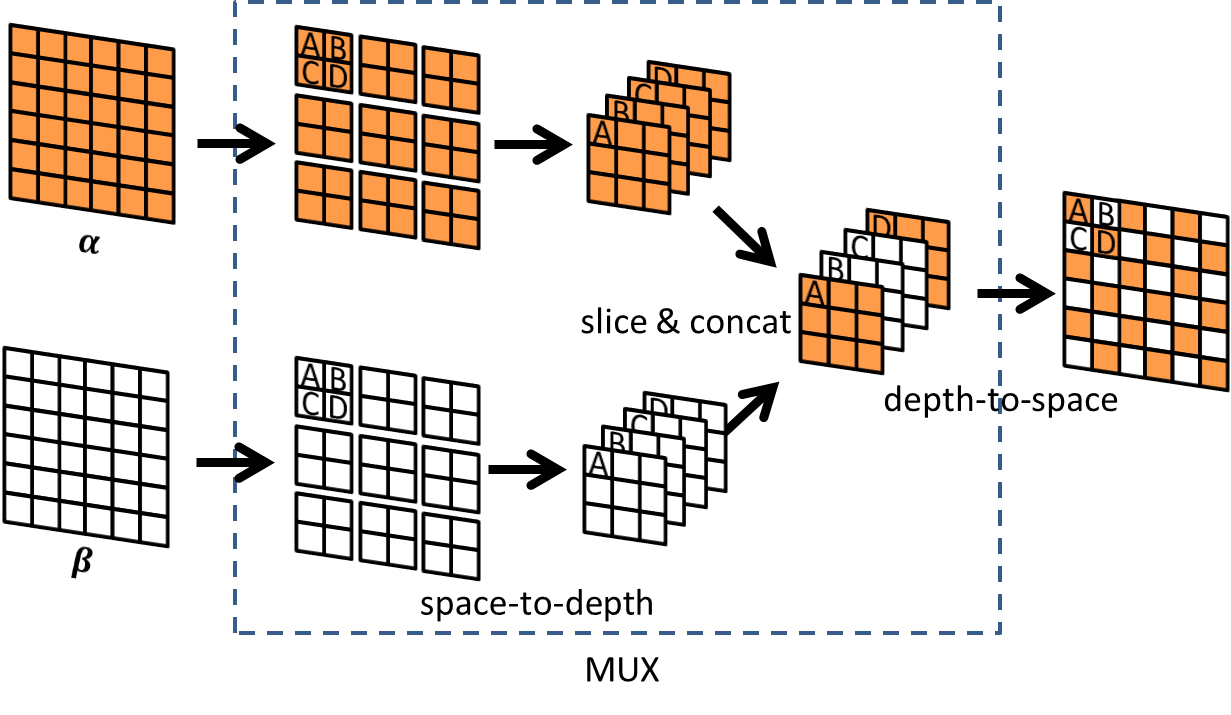}
    \caption{Multiplexer implementation with slicing and concatenating operators.}
    \label{fig:mux}
\end{figure}
\begin{figure*}
    \centering
    \includegraphics[height=4cm]{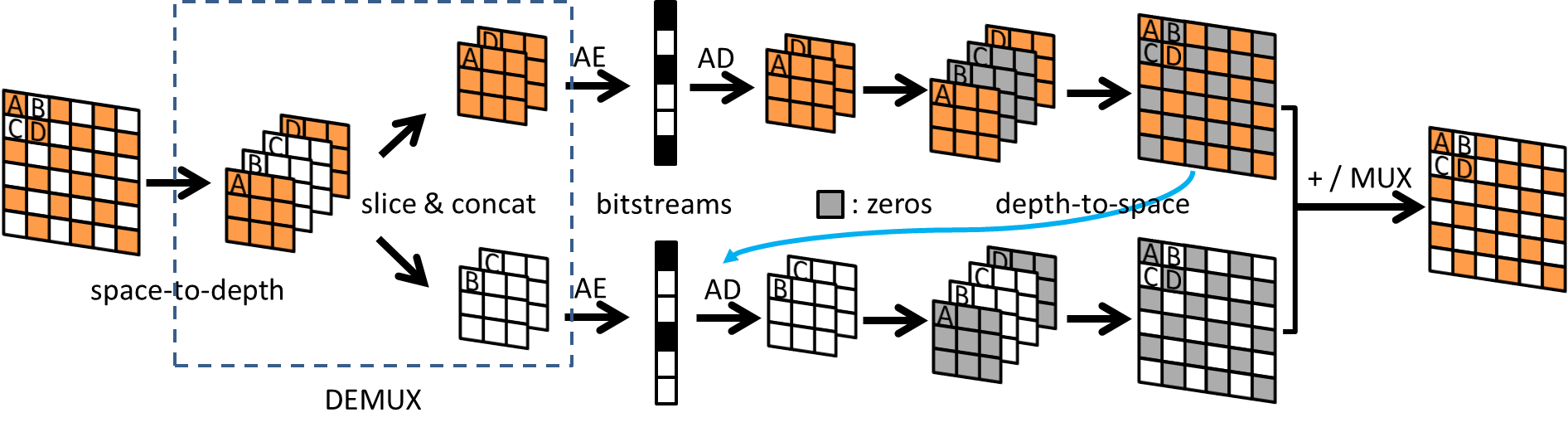}
    \caption{Demultiplexer (DEMUX) implementation with slicing and concatenating operators. Anchors and non-anchors are decoded from bitstream in turn and finally get gathered.}
    \label{fig:demux}
\end{figure*}

For frameworks which do not support slice-assign operations, also we could mix anchors and non-anchors with shape-transforming operations like reshaping, permuting, slicing or concatenating. See Figure~\ref{fig:mux}. Firstly, we flatten each $2\times 2$ patches using a space-to-depth operator (or a series of permuting and reshaping) and then slice every feature maps, anchor and non-anchor, into four chunks and concatenate four of eight chunks to get what we require. Since most popular frameworks support above mentioned operators, this provides a more general solution to implement the proposed checkerboard model.

\subsection{Demultiplexer: Split Latents for Encoding}

During encoding, latents must get divided into two chunks: anchors and non-anchors, and then be encoded by AE separately. Otherwise, future decoding will fail because AD must read all anchors from the bitstream before decoding non-anchors. So we also require a cheap operation to separate latents into anchors and non-anchors. We use an inverse operation of slice-and-concatenate multiplexer introduced in Section~\ref{sec:using-slice-and-concat} and Figure~\ref{fig:mux}. See Figure~\ref{fig:demux}, this new operation called demultiplexer performs a space-to-depth and then slices the latents or entropy parameters along channels to get anchors and non-anchors (or their corresponding entropy parameters). 

\subsection{Merging Two Chunks During Decoding}
Figure~\ref{fig:demux} also shows how to decode and recover the two chunks of latents from the bitstream (right part in the diagram). With it we further discuss how we implement the two-pass decoding. After viewing, filler chunks will get inserted into the decoded feature maps via slicing and concatenating (in our implementation, we simply use zero tensors as fillers). Then we reshape the filled feature maps to get $\vhaty_{\mathrm{anchor}}$ or $\vhaty_{\mathrm{non-anchor}}$ with sizes of $H\times W\times M$, as is described in Figure~5 in the main body. After finally decoding all anchors and non-anchors, we merge the two chunks with above mentioned multiplexer (or an addition operation if using zero fillers).

\begin{figure*}
    \centering
    \subfigure[ground truth]{
        \includegraphics[width=5cm]{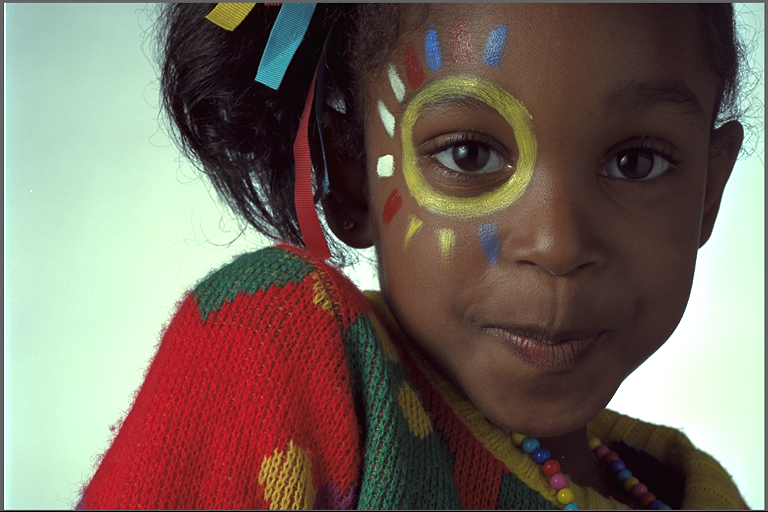}
    }
    \subfigure[serial (BPP = 0.219, PSNR = 31.96)]{
        \includegraphics[width=5cm]{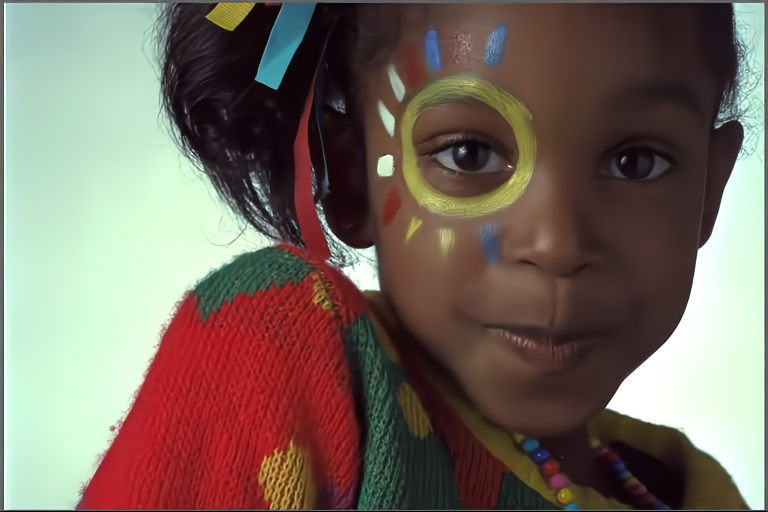}
    }
    \subfigure[parallel (BPP = 0.227, PSNR = 31.80)]{
        \includegraphics[width=5cm]{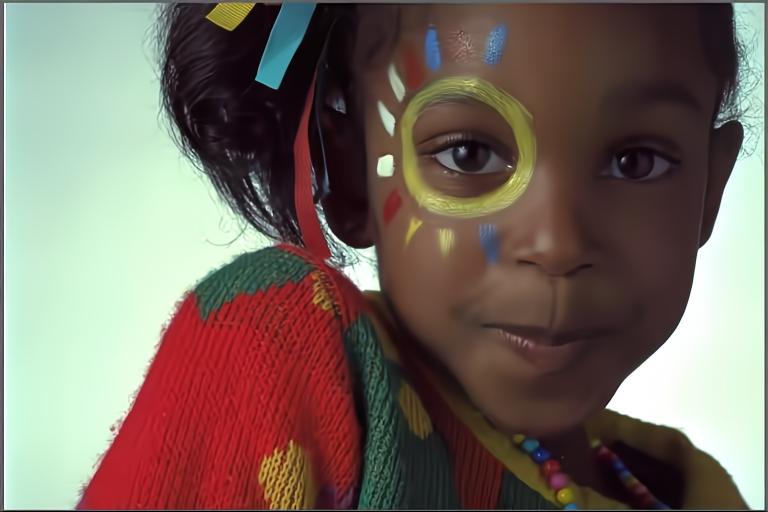}
    }
    
    \subfigure[ground truth]{
        \includegraphics[width=5cm]{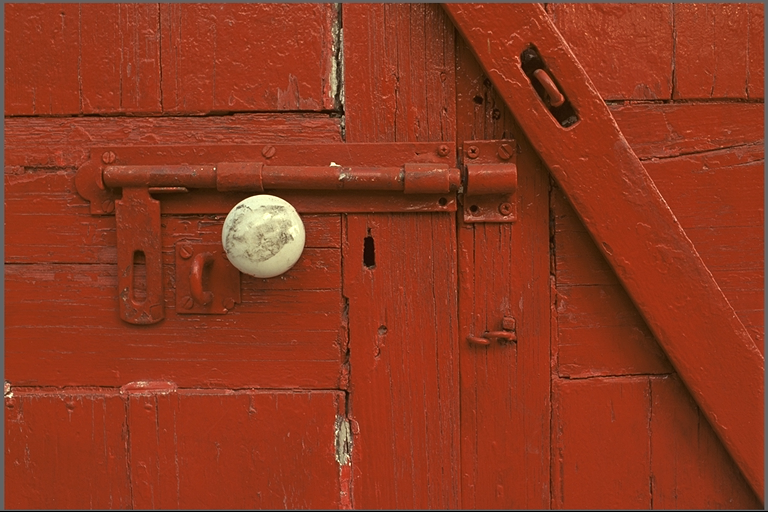}
    }
    \subfigure[serial (BPP = 0.449, PSNR = 33.54)]{
        \includegraphics[width=5cm]{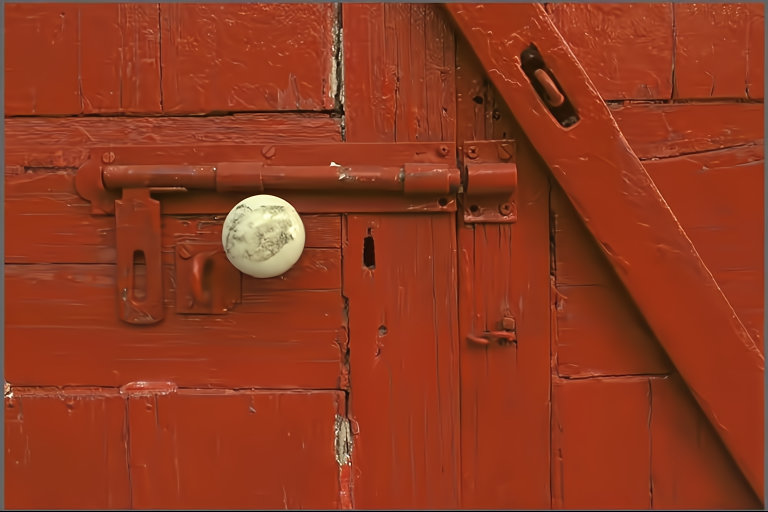}
    }
    \subfigure[parallel (BPP = 0.454, PSNR = 33.42)]{
        \includegraphics[width=5cm]{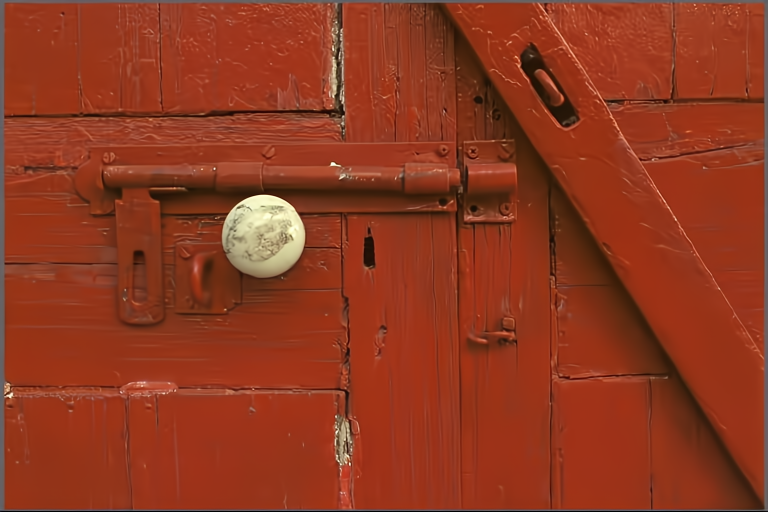}
    }
    
    \subfigure[ground truth]{
        \includegraphics[width=5cm]{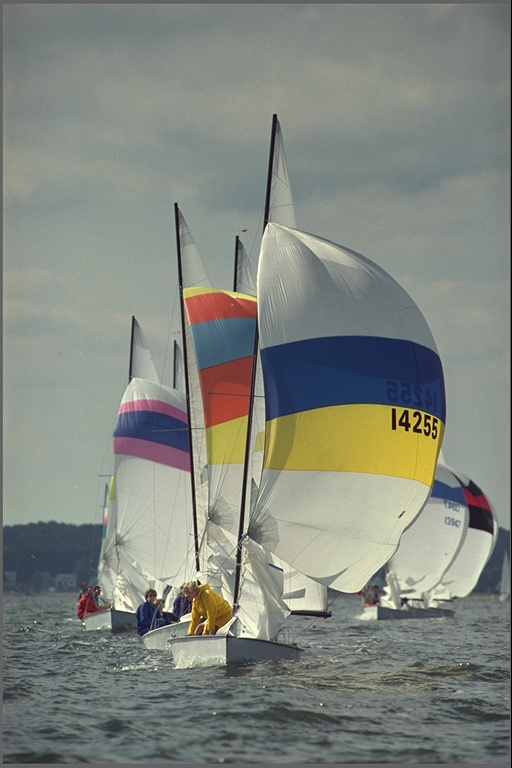}
    }
    \subfigure[serial (BPP = 0.091, PSNR = 28.29)]{
        \includegraphics[width=5cm]{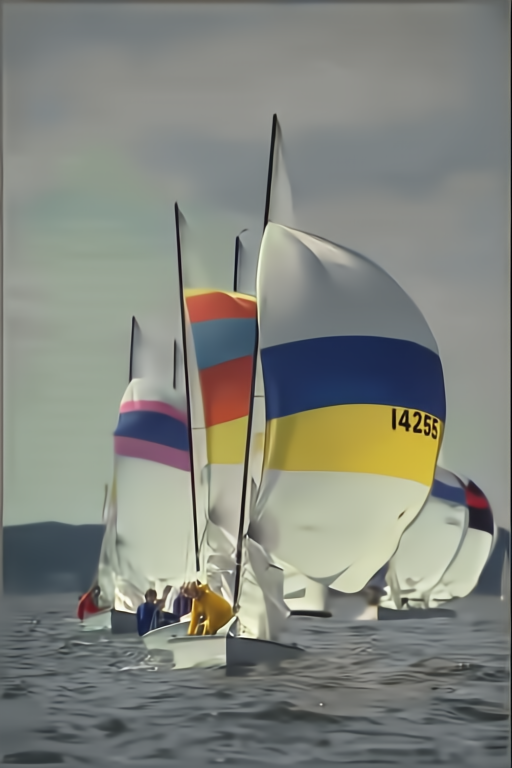}
    }
    \subfigure[parallel (BPP = 0.095, PSNR = 28.01)]{
        \includegraphics[width=5cm]{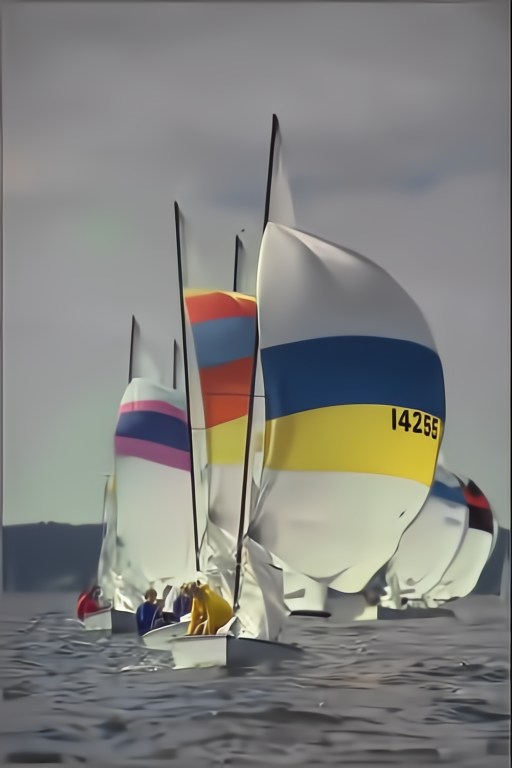}
    }
    \caption{Reconstructed images \textit{kodim15}, \textit{kodim02} and \textit{kodim09} from Kodak. The \textit{serial} ones are evaluated on our reproduced Cheng2020 models optimized for MSE and the \textit{parallel} ones are from Cheng2020 with proposed checkerboard context model optimized for MSE.}
    \label{fig:qualitative}
\end{figure*}

\begin{figure*}
    \centering
    \includegraphics[width=16cm]{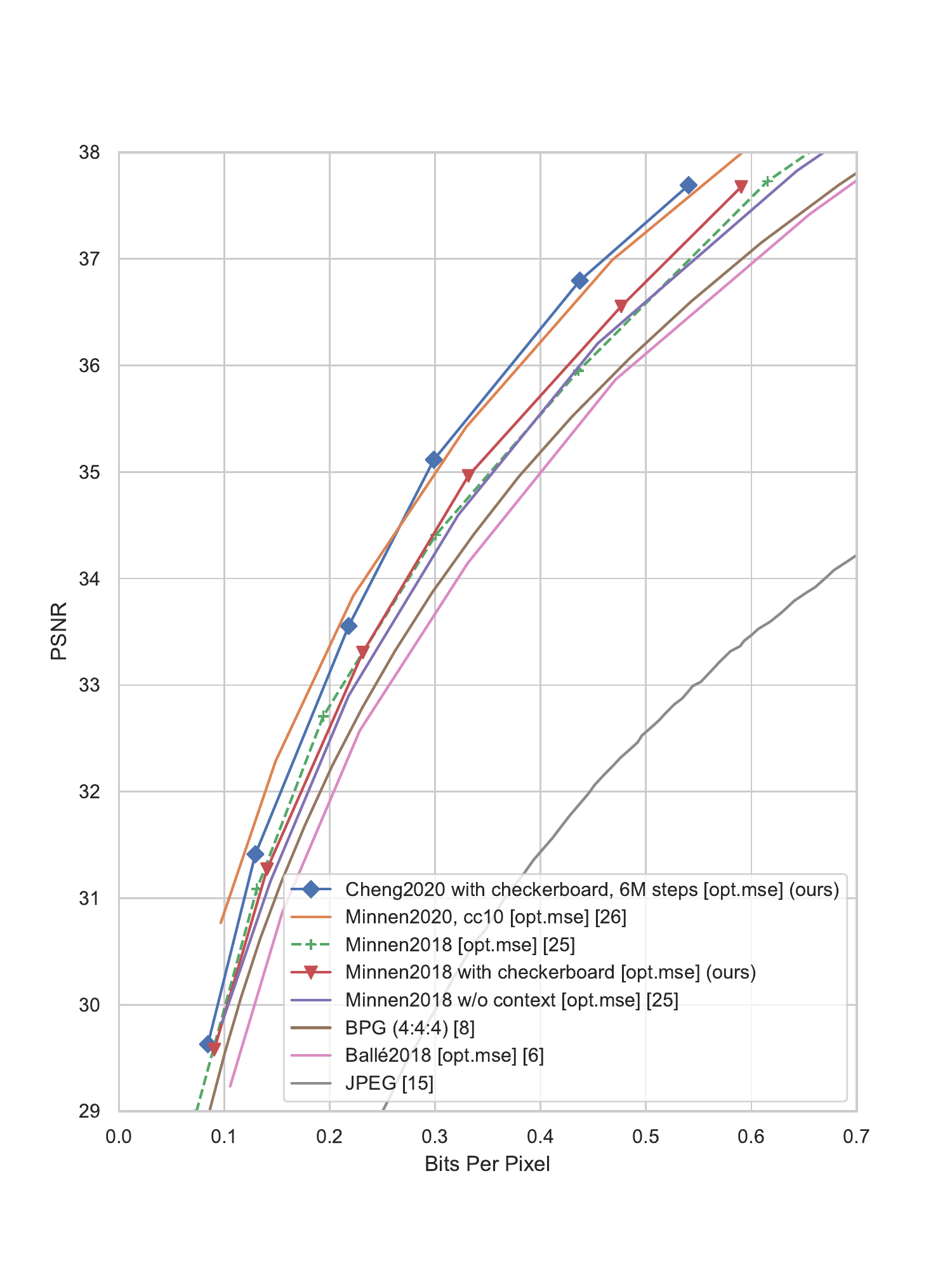}
    \caption{PSNR evaluation on Tecnick. Dash lines represent models adopting the serial context model.}
    \label{fig:rd-tecnick}
\end{figure*}

\begin{figure*}
    \centering
    \includegraphics[width=16cm]{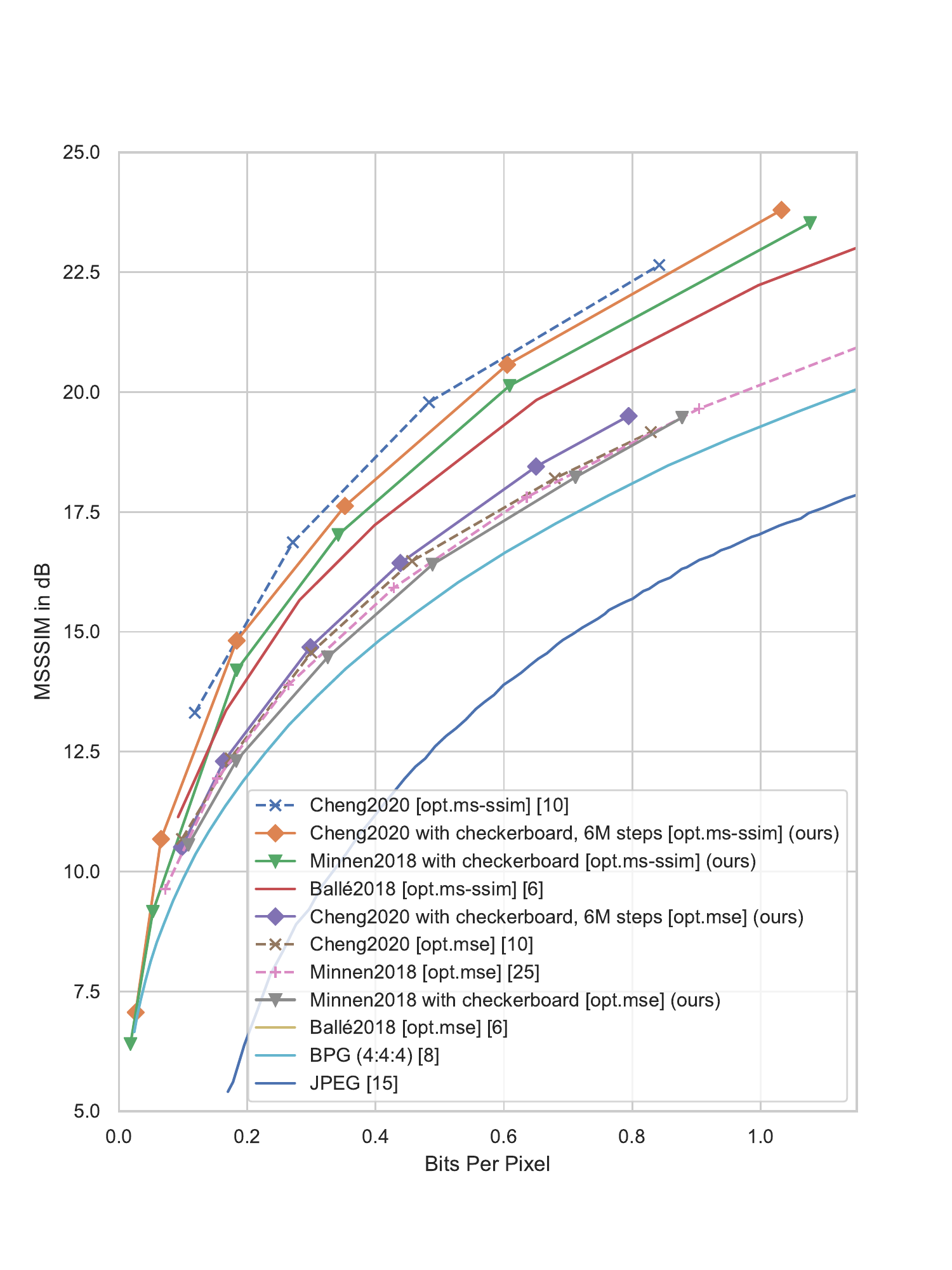}
    \caption{MS-SSIM evaluation on Kodak. Results are converted to decibels: $-10\log_{10}(1-\mathrm{MSSSIM}(x, \hat x))$. Dash lines represent models adopting the serial context model.}
    \label{fig:rd-ms}
\end{figure*}